\documentclass[%
floatfix,
 aip,
 jcp,
amsmath,amssymb,
reprint,%
]{revtex4-1}

\usepackage{graphicx}
\usepackage{dcolumn}
\usepackage{bm}
\usepackage{multirow}

\usepackage[utf8]{inputenc}
\usepackage[T1]{fontenc}
\usepackage{mathptmx}
\usepackage{etoolbox}
\usepackage{physics}
\usepackage{siunitx}
\usepackage{mathtools}
\usepackage{titlesec}
\usepackage{upgreek}
\usepackage{xr}

\makeatletter
\def\@email#1#2{%
 \endgroup
 \patchcmd{\titleblock@produce}
 {\frontmatter@RRAPformat}
 {\frontmatter@RRAPformat{\produce@RRAP{*#1\href{mailto:#2}{#2}}}\frontmatter@RRAPformat}
 {}{}
}%
\makeatother
\begin{document}
\titlespacing\section{5pt}{8pt plus 2pt minus 2pt}{5pt plus 2pt minus 2pt}
\titlespacing\subsection{5pt}{8pt plus 2pt minus 2pt}{5pt plus 2pt minus 2pt}
\titlespacing\subsubsection{5pt}{8pt plus 2pt minus 2pt}{5pt plus 2pt minus 2pt}

\title{Complex-variable MP2 theory applied to core-vacant states for the 
computation of Auger spectra}
\author{Florian Matz}
\author{Jan Philipp Drennhaus}
\author{Anthuan Ferino-Pérez}
\author{Thomas-C. Jagau}%
\email{thomas.jagau@kuleuven.be}
\email{florian.matz@kuleuven.be}
\affiliation{Department of Chemistry, KU Leuven, B-3001 Leuven, Belgium}

\date{\today}


\begin{abstract}
We model Auger spectra using second-order M\o ller-Plesset perturbation (MP2) 
theory combined with complex-scaled basis functions. For this purpose, we 
decompose the complex MP2 energy of the core-hole state into contributions from 
specific decay channels and propose a corresponding equation-of-motion (EOM) 
method for computing the doubly ionized final states of Auger decay. These 
methods lead to significant savings in computational cost compared to our 
recently developed approaches based on coupled-cluster theory [F. Matz and T.-C. Jagau, J. Chem. Phys. 
156, 114117 (2022)].

The test set for this study comprises water, ammonia, methane, hydrogen 
sulfide, phosphine, and silane. The energies of the final states of Auger 
decay are obtained with an accuracy comparable to EOM coupled-cluster 
singles and doubles (CCSD) theory. Partial decay widths and branching 
ratios between KLL, KLM, and KMM decay of K-shell holes in third-row 
hydrides are in good agreement with EOM-CCSD, while deviations are more 
significant for second-row hydrides. For L$_1$-shell holes, which undergo 
Coster-Kronig decay, MP2 yields unphysical results. However, we show that 
a suitable shift of the MP2 energy denominators leads to more reliable 
branching ratios and spectra for these problematic cases. 
\end{abstract}

\maketitle


\section{Introduction}
Auger decay~\cite{auger23} is a relaxation process of core-vacant states 
in which a less strongly bound electron refills the core vacancy while a second 
electron, the Auger electron, is emitted. It is the dominating decay 
process for core vacancies in molecules composed of light nuclei. 
Core-vacant states are produced under x-ray irradiation,\cite{auger23,
rennie00,carniato20} by collisions with high-energy electrons,\cite{spohr70} 
or by electron capture by unstable nuclei.\cite{loveland17} The measurement 
of the energy distribution of Auger electrons, i. e. the recording of Auger spectra, provides a variety of chemical information about molecules,\cite{rye84,
bolognesi12,agarwal13,mcfarland14,ramasesha16,nisoli17,marchenko18,
norman18,kraus18,plekan20} clusters,\cite{tchaplyguine03} and 
nanostructures.\cite{chao07,raman11} Auger electrons are also relevant 
for radiomedicine.\cite{ku19,pirovano21}

While Auger decay can occur in atoms as light as lithium, beryllium, 
and boron\cite{hanke85}, spectroscopic and radiotherapeutic interest 
lies in heavier nuclei with many possible final electronic states in 
a large range of energies.\cite{agarwal13,kraus18,ku19,li19,pirovano21} 
Furthermore, Auger decay not only takes place in small isolated 
molecules but is also often studied in large molecules, species embedded 
in solvents or matrices, or in solids.\cite{chao07,raman11,hofmann12,
orvis19} 

It is common practice to use the x-ray notation to name Auger spectra 
according to the shells of the involved electrons. A KLL spectrum, for 
example, includes channels where a vacancy in the K-shell (1s) gets 
filled by an electron from the L-shell (2s and 2p), while a second 
electron from the same shell is emitted. The L-shell and the M-shell 
are split into subshells: the lowest orbital in each shell is labeled 
by the index 1 (L$_1$, M$_1$), and the higher-lying three orbitals by the 
index 2,3 (L$_{2,3}$, M$_{2,3}$).

Computational modeling of Auger decay is often a necessary supplement 
to experiments because it enables the definitive assignment of signals to 
channels and electron configurations.\cite{fransson16,norman18,
kraus18} An important feature of Auger decay is that it requires 
special quantum-mechanical methods to handle the coupling to the 
continuum.\cite{moiseyev11,matz22,jagau22} 

One can distinguish between methods that only aim at the energy of 
the Auger electrons and methods where the decay channels' intensities 
are explicitly calculated, allowing for the prediction of peak heights 
and shapes in the spectrum. To account for the decay and compute 
partial widths for the decay channels, different methods have been 
proposed. One approach is Fano's theory,\cite{fano61,feshbach62,
lowdin62,averbukh05,inhester12,inhester14,kolorenc20,skomorowski21a,
skomorowski21b,gorczyca00,garcia09} which either requires an explicit 
treatment of the emitted electron or a treatment in terms of Stieltjes 
imaging.\cite{carravetta87} Often, the core-valence separation (CVS) 
is applied to the core-vacant state.\cite{cederbaum80,coriani15,
vidal19} A possible approach to facilitating calculations based 
on Fano's theory is the one-center approximation.\cite{siegbahn75,
grell20,gerlach22,tenorio22}


Recently, a different method has been introduced for modeling Auger 
decay,\cite{matz22,matz23a,matz23b,jayadev23,drennhaus24,
ferinoperez24} where the outgoing electron is described implicitly 
through complex scaling of the coordinates in the Hamiltonian\cite{
aguilar71,balslev71,aguilar71,balslev71,moiseyev11} or the basis 
functions.\cite{mccurdy78,moiseyev79} This scaling has the effect 
that the wave functions of decaying states become $L^2$ integrable 
and can be treated by standard quantum-chemical methods.\cite{
bravaya13,jagau14,zuev14,white15a,white15b,white17,jagau17} The 
eigenenergies of the Hamiltonian become complex, and their imaginary part is related 
to the decay width of the respective state, which is inversely 
proportional to the lifetime. The partial widths can be computed 
via decomposition of the energy\cite{matz22} or by restricting 
the excitation manifold by means of Auger Channel Projectors (ACPs).\cite{matz23a}

The current approaches to Auger decay based on complex-variable 
methods use coupled-cluster theory with single and double substitutions 
(CCSD). The largest systems to which we have applied these methods so 
far are the benzene molecule\cite{jayadev23} and the zinc atom.\cite{ferinoperez24} 
This limited scope is a consequence of the large basis sets that 
need to be employed in complex-variable calculations.\cite{matz22} 
It is evident that a more economical method is necessary for the 
description of Auger decay in larger molecules. Recently, we 
investigated the performance of the configuration interaction 
singles (CIS) method for the description of Auger decay.\cite{matz23a} 
This method yields reliable results, but a drawback is the need 
to run calculations based on several different reference states, 
which is inconvenient and can lead to internal inconsistencies.

In this work, we explore the performance of complex-variable 
second-order M{\o}ller-Plesset perturbation (MP2) theory for the 
description of Auger decay in several hydrides of second-row and 
third-row elements. In third-row elements, core holes can be created 
in the K-shell and in the L-shell, and the decay can involve electrons 
from the L- and M-shells, giving rise to several energetically separate 
branches of the spectrum for each core hole.

The MP2 method is well established for the computation of core ionization 
energies.\cite{triguero98,duflot10,shim11,ljubic14,kovac14,su16,smiga18} 
In a recent study, some of us combined MP2 with complex-scaled basis 
functions to describe Auger decay in the hexaaquazinc(II) complex.\cite{
ferinoperez24} In the present work, we provide a more comprehensive 
assessment of the method.

In addition to the computation of total and partial Auger decay 
widths with complex-variable MP2 theory, we also propose a 
second-order perturbative approximation to the equation-of-motion 
(EOM) double ionization potential (DIP) CCSD method [EOMDIP-CCSD(2)] 
for computing the final states of Auger decay. Compared to 
EOMDIP-CCSD,\cite{nooijen97,sattelmeyer03} which is established 
for the computation of double ionization energies, this method 
circumvents the need to carry out a CCSD calculation on the 
reference state so that the overall calculation is less costly.

The remainder of this article is structured as follows: In Section 
\ref{sec:th}, a brief overview of the theoretical background and an 
explicit expression for the partial widths in MP2 theory are given. 
We also discuss why MP2 yields unphysical widths for several decay 
processes and how this problem can be fixed by a shift of the energy 
denominator. In Section \ref{sec:compd}, the computational details are 
discussed, before results for K-edge ionization energies, decay widths, 
and Auger spectra of water, ammonia, and methane are presented in Section 
\ref{sec:res1} and of hydrogen sulfide, phosphine, and silane in Section 
\ref{sec:res2}. Section \ref{sec:res3} discusses results for L$_1$-edge 
ionization of third-row hydrides. This article ends with our general 
conclusions in Section \ref{sec:con}.

\section{Theoretical considerations} \label{sec:th}
The modeling of Auger spectra requires the computation of partial decay widths for 
peak intensities and Auger electron energies for peak positions. The latter 
are obtained from the energy differences between the initial core-ionized 
and final doubly ionized states. We discuss the computation of widths and 
initial-state energies in Secs.\,\ref{sec:th1} and \ref{sec:thmp2} and that 
of final-state energies in Sec.\,\ref{sec:th2}.

\subsection{Computation of Auger decay widths with complex scaling} \label{sec:th1}
We compute decay widths by including complex-scaled functions of the form
\begin{equation} \label{eq:bas}
\chi(r) = P(r) \, \text{exp} \big[ {-\alpha \, e^{-2i\theta}\, (r-r_A)^2} \big]
\end{equation}
in a Gaussian basis set.\cite{mccurdy78,white15a} In Eq.\,\eqref{eq:bas}, 
$P(r)$ is a polynomial in the spatial coordinates, $A$ is a nucleus, and 
$\theta$ is the complex scaling angle. The choice of exponents $\alpha$ 
for the complex-scaled functions has been extensively discussed in our 
previous works.\cite{matz22,matz23a,drennhaus24,ferinoperez24} Details 
of the basis set employed in the present work are given in Section 
\ref{sec:compd}.

In such a basis set, the molecular electronic Hamiltonian has eigenstates 
with complex energy
\begin{equation} \label{eq:cxe}
E_\text{res} = E_R - i \, \Gamma/2
\end{equation}
that correspond to resonances with energy $E_R$ that decay with the 
width $\Gamma$ corresponding to the lifetime $1/\Gamma$.\cite{siegert39,
moiseyev11} The method of complex basis functions (CBFs) is applicable 
not only to Auger decay but also to temporary anions,\cite{white15a,
white15b,white17,camps25} superexcited Rydberg states,\cite{creutzberg23} 
interatomic Coulombic decay,\cite{parravicini23} and ionization in static 
electric fields.\cite{jagau18,hernandez19}

Our previous work on Auger decay\cite{matz22,matz23a} was based on
coupled-cluster theory.\cite{bartlett12,sneskov12} Here, the wave function 
is expressed as
\begin{equation}
|\Psi_\text{CC} \rangle = e^{\hat{T}} |\Psi_0 \rangle = (1 + \hat{T} + 
\hat{T}^2/{2!} + \hat{T}^3/{3!} + \dots) |\Psi_0\rangle~.
\end{equation}
Within the CCSD approximation, the cluster operator only includes single 
and double excitations, 
\begin{equation}
\hat{T} = \hat{T}_1+\hat{T}_2 = \sum_i^{n_\text{occ}} \sum_a^{n_\text{vrt}} 
t_i^a \, a^\dagger i + \frac{1}{4} \sum_{ij}^{n_\text{occ}} 
\sum_{ab}^{n_\text{vrt}} t_{ij}^{ab} \, a^\dagger i b^\dagger j,
\end{equation} 
with $i, j, \dots$ and $a, b, \dots$ referring to occupied and virtual 
spin orbitals, respectively. For a CCSD wave function built on top of 
a core-ionized Hartree-Fock (HF) wave function $|\Psi_0\rangle$ in a 
basis set with complex-scaled functions, we showed that Auger decay is 
represented by double excitations where $i,j$ refer to the valence orbitals 
and $a$ or $b$ is the core-ionized orbital. The CC energy 
\begin{equation} \label{eq:ecc}
E_\text{CC} = E_\text{HF} + \sum_{ij}^{n_\text{occ}} \sum_{ab}^{n_\text{vrt}} 
\Big( \frac{1}{4} t_{ij}^{ab} + \frac{1}{2} t_i^a t_j^b \Big) 
\langle ij || ab \rangle
\end{equation}
computed from such a wave function is complex. 

A complication arises from the dependence of the complex energy on the scaling 
angle $\theta$. While the energy is independent of $\theta$ in a complete basis 
set,\cite{jagau22} it needs to be optimized for every state in a finite basis 
set. As suggested in previous works,\cite{moiseyev78,moiseyev79} we do this by 
minimizing $|\text{d}(E^\text{IP}_\text{CC} - E^0_\text{CC})/\text{d}\theta|$, where 
$E_\text{CC}^\text{IP}$ and $E_\text{CC}^0$ are the CC energies of the core-ionized 
state and the neutral ground state, respectively. The values obtained for $\theta$ in the 
present work are available from the supplementary material. The total Auger 
decay width $\Gamma$ is then evaluated at the optimal $\theta$ according to 
Eq.\,\eqref{eq:cxe} as 
\begin{equation} \label{eq:gamma}
\Gamma = -2\, \text{Im}(E_\text{CC}^\text{IP} - E_\text{CC}^0).
\end{equation}
As shown in our previous work,\cite{matz22} taking into account the imaginary 
part of the energy of the neutral ground state, which vanishes in the complete 
basis set limit, leads to markedly better results. 

The partial width $\gamma_{ij}(\mathbf{c})$ for decay into a particular channel 
defined by spin orbitals $i$ and $j$ can be computed as\cite{matz22} 
\begin{equation} \label{eq:gcc}
\gamma_{ij}^\text{CCSD} (\mathbf{c}) = -2 \, \text{Im} \Big( \sum_a^{n_\text{vrt}} 
\big( 2\, t_i^a t_j^\mathbf{c} + t_{ij}^{a\mathbf{c}} \big) 
\langle ij || a\textbf{c} \rangle \Big),
\end{equation}
where we have denoted the core orbital that is vacant in the initial state 
by a bold $\mathbf{c}$. It should be noted that these spin-orbital 
pairs are zeroth-order approximations to the physical decay channels. 
However, to be consistent with our previous works, we use the term 
``channel'' for them.


\subsection{Combination with MP2 theory} \label{sec:thmp2}

In this work, we extend our approach to MP2 theory. This is possible because the first-order wave 
function includes the doubly excited configurations that describe Auger 
decay when a core-ionized HF state is used as reference wave function. 
For a canonical HF reference, the MP2 energy reads
\begin{equation} \label{eq:emp2}
E_\text{MP2} = E_\text{HF} - \frac{1}{4} \sum_{ij}^{n_\text{occ}} \sum_{ab}^{n_\text{vrt}} 
\frac{\langle ab || ij \rangle \langle ij || ab \rangle}{\varepsilon_a + 
\varepsilon_b - \varepsilon_i - \varepsilon_j} = E_\text{HF} + \frac{1}{4} 
\sum_{ij}^{n_\text{occ}} \sum_{ab}^{n_\text{vrt}} t_{ij}^{ab} \, 
\langle ij || ab \rangle,
\end{equation}
with $\varepsilon_i$, $\varepsilon_j$ and $\varepsilon_a$, $\varepsilon_b$ as the 
energies of the occupied and virtual HF orbitals, respectively.

Eq.\,\eqref{eq:emp2} is identical to Eq.\,\eqref{eq:ecc} except for the absence 
of single excitations. From this, it follows that the total decay width can be 
computed from the imaginary part of the difference between the neutral-state 
and the ionized-state MP2 energies, in analogy to Eq.\,\eqref{eq:gamma}, while 
partial widths can be evaluated as 
\begin{equation}\label{eq:gmp2}
\gamma_{ij}^{\text{MP2}} (\mathbf{c}) = -2\, \text{Im} \Big( \sum_a^{n_\text{vrt}} 
\frac{\langle a\mathbf{c} || ij \rangle \langle ij || a\mathbf{c} \rangle}{\varepsilon_a 
+ \varepsilon_\mathbf{c} - \varepsilon_i - \varepsilon_j} \Big)~.
\end{equation}

In our CC-based work on Auger decay, we distinguished the approach for partial 
widths based on Eq.\,\eqref{eq:gcc} from an alternative method, dubbed Auger 
Channel Projector (ACP),\cite{matz23a} in which all excitations $a^\dagger 
\mathbf{c}^\dagger i j$ contributing to a particular partial width $\gamma_{ij}$ 
are excluded from the excitation manifold while the CC equations are solved. 
These two approaches become identical for the MP2 method since the 
amplitudes are not determined iteratively and, thus, do not depend on each 
other. In fact, partial widths can be computed from Eq.\,\eqref{eq:gmp2} 
without carrying out a full MP2 calculation: for describing the decay 
channel associated with orbitals $i$ and $j$, only the integrals 
$\langle ij || a\mathbf{c} \rangle$ need to be computed. Overall, the 
computational cost of the determination of the decay widths is decreased 
from iterative $\mathcal{O}(N^6)$ for CCSD to non-iterative
$\mathcal{O}(N^5)$ for MP2. 

\begin{table} \centering
\caption{Comparison of energies relevant for selected Auger decay channels 
of various molecules as computed at the HF and EOM-CCSD levels of theory 
using the cc-pCVTZ (5sp) basis set augmented by additional complex-scaled shells.$^a$ 
All values in eV.}
\setlength{\tabcolsep}{6pt}
\begin{tabular}{lrrrrr} \hline
 & H$_2$S/L$_1$ & PH$_3$/L$_1$ & SiH$_4$/L$_1$ & H$_2$O/K & H$_2$S/K \\ \hline
 & \multicolumn{5}{c}{HF orbital energies,$^b$ ionized state} \\ \hline 
$\varepsilon_\text{core}$ & --224.9 & --186.5 & --151.5 & --518.1 & --2440.8 \\
$\varepsilon_1$ & --204.5 & --168.4 & --135.8 & --51.7 & --282.2 \\
$\varepsilon_2$ & --22.5 & --21.2 & --22.2 & --33.8 & --231.7 \\
$\varepsilon_1 + \varepsilon_2$ & --227.0 & --189.6 & --158.0 & --85.5 & --513.9 \\
$E_\text{Auger}^c$ & --2.1 & --3.1 & --6.5 & 432.6 & 1926.9 \\ \hline
 & \multicolumn{5}{c}{HF orbital energies,$^b$ neutral state} \\ \hline
$\varepsilon_\text{core}$ & --244.0 & --203.7 & --166.5 & --559.6 & --2502.5 \\
$\varepsilon_1$ & --180.9 & --146.4 & --115.0 & --36.7 & --244.0 \\ 
$\varepsilon_2$ & --10.5 & --10.6 & --13.3 & --15.9 & --180.9 \\
$\varepsilon_1+\varepsilon_2$ & --191.4 & --157.0 & --128.3 & --52.6 & --424.9 \\
$E_\text{Auger}^c$ & 52.6 & 46.7 & 38.2 & 507.0 & 2077.6 \\ \hline
 & \multicolumn{5}{c}{EOM-CCSD energies} \\ \hline
$E_\text{IP}$ & 235.0 & 195.4 & 159.0 & 541.4 & 2475.4 \\
$E_\text{DIP}$ & 197.9 & 162.1 & 132.3$^e$ & 67.5 & 456.2$^f$ \\
$E_\text{Auger}^d$ & 37.1 & 33.3 & 26.7 & 473.9 & 2019.2 \\ \hline
\end{tabular} \raggedright

$^a$ 2 complex-scaled s-, p-, and d-shells for K-edge decay,\\
and 5 complex-scaled s-, p-, and d-shells for L$_1$-edge decay. \\
For details, see Sec.\,\ref{sec:compd}. \\
$^b$ Relevant orbitals are 
$\varphi_\text{core}$ = 2a$_1^\beta$ (H$_2$S, PH$_3$, SiH$_4$/L$_1$), \\ 
1a$_1^\beta$ (H$_2$O, H$_2$S/K); 
$\varphi_1$ = 1b$_1^\beta$ (H$_2$S/L$_1$), 3a$_1^\beta$ (PH$_3$/L$_1$), \\
1t$_2^\beta$ (SiH$_4$/L$_1$), 2a$_1^\beta$ (H$_2$O, H$_2$S/K); 
$\varphi_2$ = 2b$_1^\alpha$ (H$_2$S/L$_1$), \\
5a$_1^\alpha$ (PH$_3$/L$_1$), 2t$_2^\alpha$ (SiH$_4$/L$_1$), 
3a$_1^\alpha$ (H$_2$O, H$_2$S/K). \\
$^c$ Computed as $E_\text{Auger} = \varepsilon_1 + \varepsilon_2 - 
\varepsilon_\text{core}$. \\
$^d$ Computed as $E_\text{Auger} = E_\text{IP} - E_\text{DIP}$. \\
$^e$ $^1$A$_1$ state resulting from $1$t$_2 \otimes 2$t$_2$. \\
$^f$ Obtained using extrapolation according to Eq.\,\eqref{eq:dipmod}.
\label{tab:orb}
\end{table}

As we will illustrate numerically in Secs.\,\ref{sec:res1} and 
\ref{sec:res3}, straightforward application of Eq. \eqref{eq:gmp2} 
leads to bad results in some cases. This can be understood by analyzing 
the energy differences relevant to Auger decay. In a one-electron picture 
using Koopmans' theorem, the Auger electron energy is 
\begin{equation} \label{eq:orbe}
E_\text{Auger} = \varepsilon_1 + \varepsilon_2 - \varepsilon_\text{core}
\end{equation}
for a decay channel where an initial vacancy in $\varphi_\text{core}$ 
is filled by an electron from valence orbital $\varphi_1$ and a second 
electron is ejected from $\varphi_2$.

Tab.\,\ref{tab:orb} shows these energies for some representative examples 
of K-edge decay and L$_1$-edge decay. In addition, the Auger electron 
energies from EOM-CCSD calculations are shown. Tab.\,\ref{tab:orb} illustrates 
that Auger electron energies computed from a L$_1$-ionized HF wave function 
according to Eq.\,\eqref{eq:orbe}, with $\varepsilon_\text{core}$ as the energy 
of the unoccupied core orbital, are negative. This is qualitatively, wrong 
as it implies that the respective decay channels are closed. This failure 
can be ascribed in part to an underestimation of the core ionization energy, 
but more importantly to the too high energies of the doubly ionized states. 
These are caused by reduced screening of the nuclear charges in the core-ionized 
wave function, resulting in valence orbital energies that do not approximate 
the energies of the doubly ionized states well. As a consequence, it can be 
expected that CBF-MP2 partial decay widths of L$_1$-ionized states evaluated 
using Eq.\,\eqref{eq:gmp2} are of low quality. 

Tab.\,\ref{tab:orb} demonstrates that using orbital energies from a neutral 
HF wave function in Eq.\,\eqref{eq:orbe} leads to qualitatively correct 
Auger energies. This suggests that CBF-MP2 partial widths can be improved 
by replacing the orbital energies $\varepsilon_i$, $\varepsilon_j$, and 
$\varepsilon_\mathbf{c}$ in Eq.\,\eqref{eq:gmp2} by those from HF wave 
functions for the corresponding neutral states, denoted $\varepsilon'$. 
This can be written as 
\begin{equation}\label{eq:gmp2mod}
\gamma_{ij}^{\text{MP2-mod}} (\mathbf{c}) = -2\, \text{Im} \Big( 
\sum_a^{n_\text{vrt}} \frac{\langle a\mathbf{c} || ij \rangle 
\langle ij || a\mathbf{c} \rangle}{\varepsilon_a 
+ \varepsilon'_\mathbf{c} - \varepsilon'_i - \varepsilon'_j} \Big)~,
\end{equation}
which can be viewed as a shift of the MP2 energy denominator. Notably, 
an approach where the energies $\varepsilon_a'$ of the virtual orbitals of 
the neutral state are used instead of $\varepsilon_a$ leads to worse results 
for the peak shapes and the sum of partial decay widths. This is shown in 
the supplementary material.

From Tab.\,\ref{tab:orb}, it is seen that the energies relevant for K-edge 
decay of H$_2$O and H$_2$S are also improved using orbital energies of the 
neutral HF wave function. However, the effect is less pronounced than for 
L$_1$-edge decay, and the use of orbital energies from the core-ionized 
wave function yields qualitatively correct results. We add that the 
importance of using qualitatively correct energy differences was already 
realized in earlier works on Auger decay in which the decay width was 
evaluated as a transition property.\cite{chase71,carravetta87}


\subsection{Computation of Auger electron energies} \label{sec:th2}
In our previous work, we used the EOMDIP-CCSD method\cite{sattelmeyer03,
nooijen97} to compute the energies of the final doubly ionized states. 
This method is based on applying a linear excitation operator 
$\hat{R}^\text{DIP}$ to the CCSD wave function,
\begin{equation} \label{eq:eomcc}
|\Psi_\text{EOM-CCSD}\rangle = \hat{R}^\text{DIP} |\Psi_\text{CCSD} \rangle = 
\hat{R}^\text{DIP} e^{\hat{T}} |\Psi_0 \rangle~.
\end{equation}
The operator $\hat{R}^\text{DIP}$ removes two electrons from the wave 
function and is defined as 
\begin{equation}\label{eq:rdip}
\hat{R}^\text{DIP} = \frac{1}{2} \sum_{ij}^{n_\text{occ}} r_{ij} \, j i 
+ \frac{1}{6} \sum_{ijk}^{n_\text{occ}} \sum_a^{n_\text{vrt}} 
r_{ijk}^a \, a^\dagger kji~.
\end{equation}

The computational cost of the most expensive step of the solution of the 
EOMDIP-CCSD eigenvalue equations scales as $\mathcal{O}(n_\text{occ}^3 
n_\text{vrt}^3)$, but it requires the preceding solution of the CCSD equations 
that entail $\mathcal{O}(n_\text{occ}^2 n_\text{vrt}^4)$ cost. To avoid 
the latter step, we use the EOM-CCSD(2) approach,\cite{stanton95} also 
referred to as EOM-MP2, where the CCSD amplitudes are replaced 
by MP2 amplitudes. 

The expressions necessary to solve the EOMDIP-CCSD(2) eigenvalue equations 
and determine the operator $\hat{R}^{\text{DIP}}$ are identical to those for 
EOMDIP-CCSD and are also given in the supplementary material. With this approach, 
the cost for determining the double ionization energies formally still scales 
as $\mathcal{O}(N^6)$, but since the number of occupied orbitals is much smaller 
than the number of virtual orbitals in typical complex-variable calculations, 
the reduction of the operation count from $\mathcal{O}(n_\text{occ}^2 
n_\text{vrt}^4)$ to $\mathcal{O}(n_\text{occ}^3 n_\text{vrt}^3)$ achieved 
by skipping the CCSD equations is substantial. In addition, the memory requirements 
are lowered from $n_\text{vrt}^4$ to $n_\text{vrt}^3n_\text{occ}$ because 
the EOMDIP equations do not involve integrals $\langle ab || cd \rangle$. 

With both methods, EOMDIP-CCSD and EOMDIP-CCSD(2), highly correlated states 
such as those involving holes in the 2a$_1$ orbital of H$_2$S are difficult to converge. 
For such states for which we could not converge the EOMDIP-CCSD or EOMDIP-CCSD(2) 
eigenvalue equations, the double ionization energies were extrapolated as detailed 
in the supplementary material. For this purpose, all double ionization energies 
were computed with a truncated EOMDIP operator 
\begin{equation} \label{eq:dipmod}
\hat{R}^{\text{DIP, mod}} = \frac{1}{2}\sum_{ij}^{n_\text{occ}} r_{ij}^\text{mod} ji~, 
\end{equation}
which only involves 2-hole excitations. A linear relation between the energy in 
the full excitation manifold and the energy in the reduced excitation manifold 
was assumed and used to estimate the energy of the states that did not converge 
in the full excitation manifold.

\subsection{Construction of Auger spectra} \label{sec:th3}

With the partial widths and the Auger electron energies from Sections \ref{sec:th1} 
and \ref{sec:th2}, we construct Auger spectra as follows:\cite{matz23b,jayadev23} 
At the position $E_K$ corresponding to the transition $K$ from the core-ionized 
state with vacant core orbital $\mathbf{c}$ to the doubly ionized state with 
vacant orbitals $i$ and $j$, a peak is centered with a broadening that accounts 
for the lifetime broadening and vibrational interference. In this work, 
we use Gaussian functions to model the peak shape as 
\begin{equation}\label{eq:spectrum}
I_K(E) = \sum_{ij} \gamma_{ij}(\mathbf{c}) \, r_{ij}^2 (K) \, 
\text{exp} \Big[ -\frac{4 \, \text{ln}(2)^2(E-E_K)^2}{\text{FWHM}} \Big],
\end{equation}
where FWHM is the full-width at half maximum of the Gaussian peak. As different spectra show different peak widths, we adjust the FWHM of our spectra accordingly while keeping the widths small enough to retain information about the composition of the spectral features in terms of final states.

Since $\gamma_{ij}(\mathbf{c})$ and $E_K$ are obtained from separate calculations, 
it is necessary to weigh every partial width $\gamma_{ij}(\mathbf{c})$ by 
$r_{ij}^2(K)$, which is the amplitude of the excitation $ji$ in the EOMDIP 
wave function corresponding to transition $K$. This procedure accounts for 
the change of the wave function during Auger decay.

We note that the extrapolation procedure discussed at the end of Section \ref{sec:th2}
would introduce an imbalance to the spectrum because $\hat{R}^{\text{DIP, mod}}$ 
does not include 3-hole-1-particle excitations. As a consequence, $\sum_{ij} r_{ij}^{2,\text{mod}}$ is always 1 while $\sum_{ij} r_{ij}^2$ is not. To remove this imbalance, the amplitudes obtained with the 
full operator $\hat{R}^{\text{DIP}}$ were renormalized as $r_{ij}^{2(\text{new})} 
= r_{ij}^{2(\text{old})} / \sum_{ij} r_{ij}^{2(\text{old})}$ before they were used 
in Eq.\,\eqref{eq:spectrum}.


\section{Computational details} \label{sec:compd}
The test set to evaluate the performance of our new CBF-MP2 method for the 
construction of Auger spectra comprises water, ammonia, methane, hydrogen 
sulfide, phosphine, and silane. All calculations were performed using a 
modified version of the Q-Chem program package, version 6.2.\cite{epifanovsky21}

Neutral water, ammonia, and methane all have 10 electrons, while hydrogen 
sulfide, phosphine, and silane all have 18 electrons. Water and hydrogen 
sulfide both belong to the C$_{2\text{v}}$ point group, such that the 
electron configurations are 
(1a$_1$)$^2$(2a$_1$)$^2$(1b$_2$)$^2$(3a$_1$)$^2$(1b$_1$)$^2$ for water and 
(1a$_1$)$^2$(2a$_1$)$^2$(1b$_2$)$^2$(3a$_1$)$^2$(1b$_1$)$^2$(4a$_1$)$^2$(2b$_2$)$^2$(5a$_1$)$^2$(2b$_1$)$^2$ 
for hydrogen sulfide. Ammonia and phosphine both belong to the 
C$_{3\text{v}}$ point group, such that the electron configurations 
are (1a$_1$)$^2$(2a$_1$)$^2$(1e)$^4$(3a$_1$)$^2$ for ammonia and (1a$_1$)$^2$(2a$_1$)$^2$(1e)$^4$(3a$_1$)$^2$(4a$_1$)$^2$(2e)$^4$(5a$_1$)$^2$ 
for phosphine. Methane and silane both belong to the T$_\text{d}$ point 
group, such that the electron configurations are 
(1a$_1$)$^2$(2a$_1$)$^2$(1t$_2$)$^6$ for methane and 
(1a$_1$)$^2$(2a$_1$)$^2$(1t$_2$)$^6$(3a$_1$)$^2$(2t$_2$)$^6$ for silane. 
The bond lengths and angles used in the calculations can be found in the 
supplementary material.


\subsection{Basis set details} \label{sec:bas}
We use the basis sets that we deemed optimal in our previous works on 
Auger decay using complex-variable methods.\cite{matz22,matz23a,matz23b} 
We start by combining s- and p-shells from the (aug-)cc-pCV5Z basis 
with d- and f-shells from the (aug-)cc-pCVTZ basis set. While for water, 
ammonia, and methane, the resulting cc-pCVTZ(5sp) basis set has proven 
adequate, we employ the augmented version for molecules involving 
third-row elements.

To these basis sets, several complex-scaled s-, p-, and d-shells were 
added for the CBF-CCSD and CBF-MP2 calculations; the exponents can be 
found in the supplementary material. In Sections \ref{sec:res1}--\ref{sec:res3}, 
we denote the addition of $n$ complex-scaled s-, p-, and d-shells by 
adding ``+$n$(spd)'' to the name of the basis set. EOMDIP-CCSD and 
EOMDIP-CCSD(2) calculations were carried out in the (aug-)cc-pCVTZ(5sp) 
basis sets without adding complex-scaled functions. 

For O, N, C, and H in calculations on water, ammonia, and methane, 
the exponents of the complex-scaled shells were taken from 
Ref.~\citenum{matz23a}, where we had determined them by scaling 
the exponents optimized for the neon atom\cite{matz22} by factors 
accounting for the diffuseness of the basis set.

For the calculations on hydrogen sulfide, phosphine, and silane, 
we first calculated the geometric average of the exponents in the 
respective aug-cc-pCVTZ (5sp) basis sets. This yielded $\bar{\zeta}$=5.50, 
3.02, 2.57, 2.09, and 0.418 for Ne, S, P, Si, and H, respectively. 
We then scaled the optimal exponents for Ne\cite{matz22} by a factor 
of $0.550 = 3.02/5.50$, $0.468 = 2.57/5.50$, $0.381 = 2.09/5.50$, 
and $0.0762 = 0.418/5.50$ for S, P, Si, and H, respectively, to obtain 
the exponents for two complex-scaled shells of each angular momentum. 
In some calculations, one more shell with an exponent between the first 
two was added to these two shells. In previous work,\cite{drennhaus24} 
we found that more than three complex-scaled shells were needed to 
simultaneously describe L$_1$L$_{2,3}$M Coster-Kronig decay and L$_1$MM 
Auger decay as the emitted electrons have very different energies. 
Therefore, in the present work, we add further complex-scaled shells 
in an even-tempered manner with a factor of 0.5, starting from the most 
diffuse of the initial two shells, to test the basis-set convergence. 



\section{K-shell ionization in water, ammonia, and methane} \label{sec:res1}
\subsection{Core ionization energies and total decay widths} \label{sec:res1a}

Tab.\,\ref{tab:2rowE} shows the core ionization energies for water, ammonia, 
and methane, computed as CCSD or MP2 energy differences between the 
neutral ground states and the core-ionized states.

\begin{table}[bht] \centering
\caption{Core ionization energies of water, ammonia, and methane computed 
with CBF-CCSD and CBF-MP2 in the cc-pCVTZ(5sp) basis set with different 
numbers of complex-scaled s-, p-, and d-shells added. All values in eV.}
\label{tab:2rowE}
\begin{tabular}{lS[table-format=3.2]S[table-format=3.2]S[table-format=3.2]}\hline
Method & {H$_2$O} & {NH$_3$} & {CH$_4$}\\ \hline
CBF-CCSD$^\text{a}$ & 539.69 & 405.56 & 290.83 \\
CBF-MP2$^\text{a}$ & 539.94 & 405.70 & 290.90 \\
CBF-MP2$^\text{b}$ & 539.92 & \dots & \dots \\
Expt. & {539.73(17)~\cite{wang2021calibration}} & {405.56~\cite{kryzhevoi11}}& {290.91~\cite{saethre97}} \\ \hline
\end{tabular} \\ \raggedright
$^\text{a}$ 2 complex-scaled s-, p-, and d-shells \\
$^\text{b}$ 4 complex-scaled s-, p-, and d-shells \\
\end{table}

The computed energies only differ by 0.07--0.25\,eV between the two 
methods and lie within the experimental error margin. To gauge the 
convergence of the MP2 results with respect to the number of 
complex-scaled shells, we conducted an additional calculation 
for water with 4 instead of 2 complex-scaled s-, p-, and d-shells, 
which only led to a 0.02\,eV decrease in the computed ionization 
potential.

We point out that the good agreement with experiment is at least partly 
due to a cancellation of errors: relativistic effects shift the core 
ionization energies by 0.38 eV (oxygen), 0.21 eV (nitrogen), and 0.10 eV 
(carbon).\cite{carbone19} When incorporating these values as corrections, 
the deviation of the present calculations from the experiment amounts to 
up to 0.35\,eV.

\begin{table}[tbh] \centering
\caption{Total decay widths of core-ionized water, ammonia, and methane 
computed with CBF-CCSD and CBF-MP2 in the cc-pCVTZ(5sp) basis set with 
different numbers of complex-scaled s-, p-, and d-shells added. All values 
in meV.}
\label{tab:2rowG}
\begin{tabular}{lS[table-format=3.1(1)]S[table-format=3.1]S[table-format=3.1(1)]|S[table-format=3.1]S[table-format=3.1]S[table-format=3.1]} \hline
&\multicolumn{3}{c|}{$\Gamma$} & \multicolumn{3}{c}{$\sum\gamma_{ij}$} \\ \hline
Method & {H$_2$O} & {NH$_3$} & {CH$_4$} & {H$_2$O} & {NH$_3$} & {CH$_4$} \\ \hline
CBF-CCSD$^\text{a}$ & 139.3 & 110.6 & 76.3 & 146.2 & 119.9 & 97.4 \\
CBF-MP2$^\text{a}$ & 190.2 & 169.5 & 121.7 & 182.7 & 162.7 & 125.4 \\
CBF-MP2$^\text{b}$ & 212.9 & \dots & \dots & 209.8 & \dots & \dots \\
Expt. & {160(2)~\cite{sankari03}} & \dots & {94(1)~\cite{carroll99}}&\dots&\dots&\dots \\ \hline
\end{tabular} \\ \raggedright
$^\text{a}$ 2 complex-scaled s-, p-, and d-shells.\\
$^\text{b}$ 4 complex-scaled s-, p-, and d-shells.\\
\end{table}

The total Auger decay widths computed according to Eq.\,\eqref{eq:gamma} 
are shown in Tab.\,\ref{tab:2rowG}. It is obvious that widths differ 
more substantially between the methods than energies. MP2 calculations 
yield widths that are up to 59 meV or 60\% larger than those computed 
with CCSD, with larger relative deviations for lighter atoms. These 
deviations are much larger than those observed in EOM ionization potential (IP)-CCSD or CIS 
calculations for the same elements in the same basis set.\cite{matz23a} 
Furthermore, our previous calculations almost always resulted in 
narrower decay widths compared to the experimental values, which 
can be rationalized by the presence of decay processes not accounted 
for by our model, for example, x-ray fluorescence or double 
Auger decay.\cite{carlson65} On the contrary, the MP2 method yields 
unphysically high decay widths even though fewer terms are included 
in the wave function compared to CCSD. The overestimation is only 
exacerbated by including more complex-scaled shells in the basis.

In complex-variable calculations for the description of Auger 
decay, the partial decay widths do not add up to the total decay 
width. This is discussed in our previous work~\cite{matz23a,matz23b,
drennhaus24} and, to a large degree, is a consequence of basis-set 
incompleteness. Therefore, a measure for the quality of the basis 
set is the deviation of the sum of partial widths ($\sum \gamma_{ij}$) 
from the total width computed with Eq.\,\eqref{eq:gamma}. For the MP2 
results in Tab.\,\ref{tab:2rowG}, this deviation is 4--8 meV (3\%--4\%), 
i.e., somewhat lower than the 7--21 meV (4\%--8\%) observed for CCSD. 
However, this lower discrepancy does not mean that either of the two 
MP2 numbers is a good estimate of the experimental decay width; CCSD 
results are clearly better.

Notably, the sum of partial widths is always larger than the total 
width with CCSD, but this is not the case for MP2, where the partial 
widths add up to lower numbers for water and ammonia. This can be 
traced back to numerical differences between MP2 and CCSD in the 
double excitation amplitudes that describe Auger decay. 



\subsection{Partial decay widths} \label{sec:res1b}

\begin{table}\centering
\caption{Partial decay widths of core-ionized water 
computed with CBF-CCSD and CBF-MP2 in the cc-pCVTZ(5sp) basis set with 
different numbers of complex-scaled s-, p-, and d-shells added. All values 
in meV.}
\label{tab:water-pw}
\begin{tabular}{lrrrr}\hline
Decay & CCSD & \multicolumn{2}{c}{MP2} & MP2-mod \\
channel & +2(spd) & +2(spd) & +4(spd) & +2(spd) \\ \hline
$^1$A$_1$ (2a$_1$2a$_1$) & 16.7 & 16.4 & 18.8 & 18.7\\
$^1$A$_1$ (2a$_1$3a$_1$) & 13.5 & 13.7 & 20.2 & 16.1\\
$^1$B$_1$ (2a$_1$1b$_1$) & 12.3 & 12.0 & 19.4 & 18.0\\
$^1$B$_2$ (2a$_1$1b$_2$) & 7.3 & 10.0 & 15.5 & 9.9\\
$^3$A$_1$ (2a$_1$3a$_1$) & 2.4 & 2.7 & 4.0 & 2.0\\
$^3$B$_1$ (2a$_1$1b$_1$) & 2.9 & 2.7 & 4.4 & 2.9\\
$^3$B$_2$ (2a$_1$1b$_2$) & 1.9 & 2.1 & 3.2 & 1.5\\
$^1$A$_1$ (3a$_1$3a$_1$) & 11.7 & 17.4 & 18.4 & 11.2\\
$^1$A$_1$ (1b$_1$1b$_1$) & 16.5 & 23.6 & 23.4 & 15.5\\
$^1$A$_1$ (1b$_2$1b$_2$) & 10.1 & 13.9 & 12.9 & 9.0\\
$^1$B$_1$ (3a$_1$1b$_1$) & 16.4 & 25.4 & 27.6 & 16.1\\
$^1$B$_2$ (3a$_1$1b$_2$) & 17.5 & 20.8 & 18.4 & 14.5\\
$^1$A$_2$ (1b$_1$1b$_2$) & 14.2 & 21.7 & 23.1 & 12.8\\
$^3$B$_1$ (3a$_1$1b$_1$) & 0.3 & 0.2 & 0.3 & 0.2\\
$^3$B$_2$ (3a$_1$1b$_2$) & 0.2 & 0.2 & 0.2 & 0.1\\
$^3$A$_2$ (1b$_1$1b$_2$) & 0.0 & 0.0 & 0.0 & 0.0\\\hline
Sum & 146.2 & 182.7 & 209.8 & 148.5 \\ \hline
\end{tabular}
\end{table}

Partial decay widths for H$_2$O, NH$_3$, and CH$_4$ have been computed 
according to Eqs.~\eqref{eq:gcc}, \eqref{eq:gmp2} and \eqref{eq:gmp2mod}. 
The results for water are presented in Tab.\,\ref{tab:water-pw}, while 
those for ammonia and methane can be found in the supplementary material. 
For some channels, CCSD and MP2 agree within less than 1 meV, whereas 
large deviations of more than 10 meV occur for other channels. In general, 
the agreement is better for channels involving the inner-valence 2a$_1$ 
orbital than for channels that only involve outer-valence orbitals. For 
the latter cases, the overestimation can amount to 50\% or more, which 
explains why MP2 yields too large total decay widths (see Tab.\,\ref{tab:2rowG}). 
Notably, increasing the size of the basis set for the MP2 calculation 
leads to an even larger disagreement, and channels involving inner-valence 
orbitals start to be affected as well. For triplet channels, deviations 
between CCSD and MP2 are lower than for singlet channels but this has 
little impact because triplet channels have low intensities anyway. 

For water, we also computed partial widths according to Eq.\,\eqref{eq:gmp2mod}. 
As expected, these MP2-mod results are in better agreement with CCSD for 
most channels. In particular, those channels whose intensities are overestimated 
by regular MP2 have a lower width and the sum of partial widths differs from 
the CCSD result by no more than 2 meV. The root mean square (rms) deviation 
with respect to CCSD decreases from 4 to 2 meV. However, there are some 
decay channels, e.g., $^1$B$_1$ (2a$_1$1b$_1$) (third line in Tab.\,
\ref{tab:water-pw}), for which MP2-mod deviates substantially from CCSD.  


\subsection{Auger spectra} \label{sec:res1c}

To construct Auger spectra, we computed the energies of the doubly ionized 
states with the EOMDIP-CCSD(2) and EOMDIP-CCSD methods. These results are 
reported in the supplementary material and are in excellent agreement for all 
molecules, as exemplified for water in Table~\ref{tab:h2oeom}. In this case, 
the maximum deviation between the methods amounts to 0.18 eV and the 
rms deviation to 0.08 eV. Similar to the partial decay widths, deviations 
are lower for higher-lying states, i.e., those with holes in inner-valence 
orbitals.

\begin{table} \centering
\caption{Lowest double ionization energies of water in eV computed 
with EOMDIP-CCSD and EOMDIP-CCSD(2) in the cc-pCVTZ(5sp) basis set.}
\begin{tabular}{llrr}\hline
Electronic & Leading & \multicolumn{2}{c}{Double ionization energies} \\
state & amplitude & EOMDIP-CCSD & EOMDIP-CCSD(2) \\ \hline                    
$^3$B$_1$ & 3a$_1$1b$_1$ & 41.01 & 41.15 \\
$^1$A$_1$ & 1b$_1$1b$_1$ & 42.04 & 42.22 \\
$^1$B$_1$ & 3a$_1$1b$_1$ & 43.44 & 43.58 \\
$^3$A$_2$ & 1b$_1$1b$_2$ & 45.08 & 45.15 \\
$^1$A$_1$ & 3a$_1$3a$_1$ & 46.48 & 46.58 \\
$^1$A$_2$ & 1b$_1$1b$_2$ & 46.84 & 46.92 \\
$^3$B$_2$ & 3a$_1$1b$_2$ & 46.90 & 46.93 \\
$^1$B$_2$ & 3a$_1$1b$_2$ & 48.99 & 49.03 \\
$^1$A$_1$ & 1b$_2$1b$_2$ & 53.79 & 53.81 \\
$^3$B$_1$ & 2a$_1$1b$_1$ & 59.54 & 59.62 \\
$^3$A$_1$ & 2a$_1$3a$_1$ & 61.25 & 61.29 \\
$^3$B$_2$ & 2a$_1$1b$_2$ & 65.20 & 65.19 \\
$^1$B$_1$ & 2a$_1$1b$_1$ & 66.25 & 66.33 \\
$^1$A$_1$ & 2a$_1$3a$_1$ & 67.49 & 67.56 \\
$^1$B$_2$ & 2a$_1$1b$_2$ & 72.45 & 72.45 \\
$^1$A$_1$ & 2a$_1$2a$_1$ & 85.90 & 85.88 \\ \hline
\end{tabular}
\label{tab:h2oeom}
\end{table}

\begin{figure} \centering
\includegraphics[width=\linewidth]{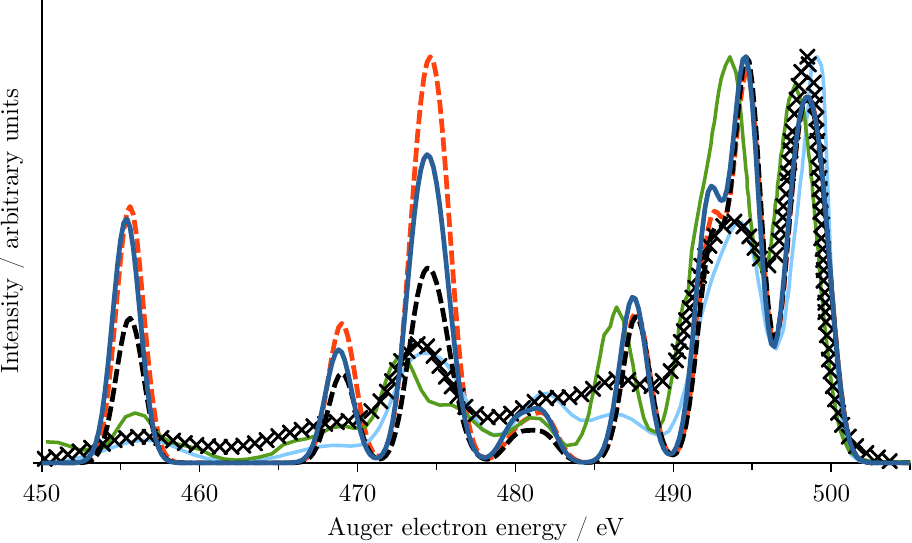}\\
\includegraphics[width=\linewidth]{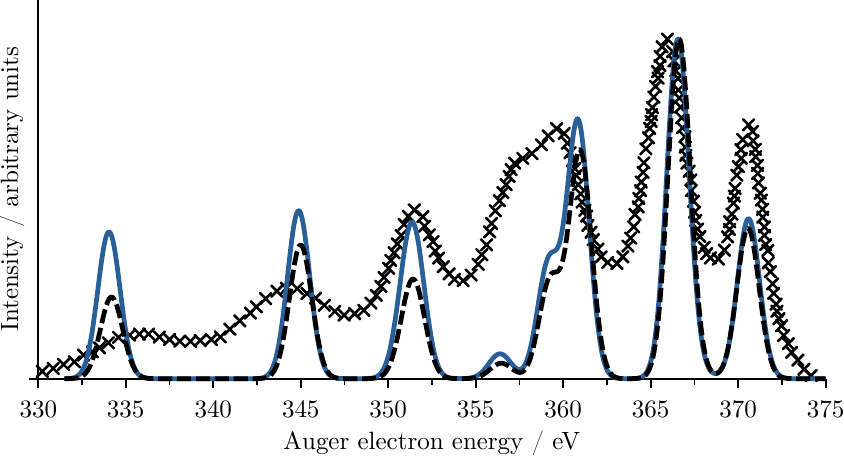}\\
\includegraphics[width=\linewidth]{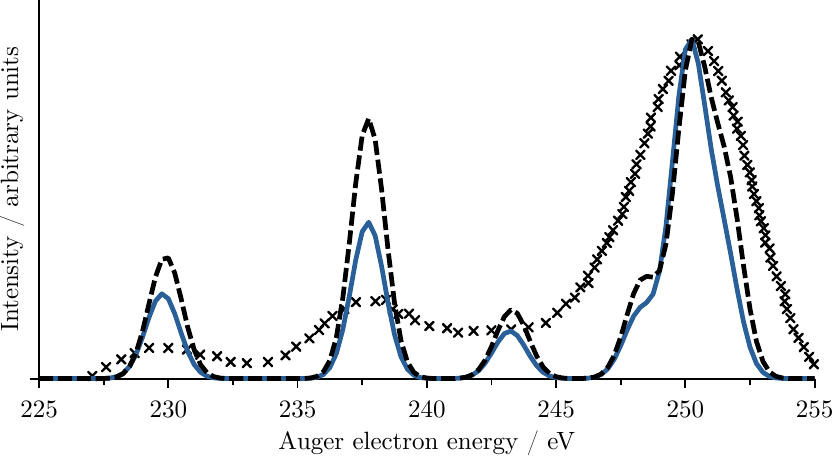}
\caption{Auger spectra of water (top), ammonia (middle), and methane 
(bottom). Dark blue full line: CBF-CCSD/EOMDIP-CCSD, black dashed line: 
CBF-MP2/EOMDIP-CCSD(2), orange dashed line: CBF-MP2-mod/EOMDIP-CCSD(2), 
green line: RASPT2/one-center approximation (Ref.~\cite{tenorio22}) light 
blue line: MRCI/single-center expansion (Ref.~\citenum{inhester12}), black $\times$: 
experiment (Refs.~\citenum{moddeman71,shaw77,kivimaki96}).}
\label{fig:KLLwater}
\end{figure}

The resulting Auger spectra are shown in Fig.\,\ref{fig:KLLwater}. 
For water and methane, our spectra are shifted by 1.7 and 1.5\,eV, 
respectively, to higher energies so that the highest-energy peak 
coincides with the experimental spectrum. The FWHM is 2.0 eV for 
water and 1.5\,eV for ammonia and methane. Consistent with the 
small deviations between EOMDIP-CCSD and EOMDIP-CCSD(2) for the 
double ionization energies, the peak positions of our theoretical 
spectra are in excellent agreement for each molecule. For water, 
we also show restricted active space second-order perturbation 
theory (RASPT2) and multireference configuration interaction 
(MRCI) spectra,\cite{inhester12,tenorio22} whose peak positions 
are in good agreement with ours. 

We normalized all spectra such that the most intense peak in every 
spectrum has the same height. This explains why the agreement between 
CBF-CCSD and CBF-MP2 is better at high Auger electron energies, 
even though the partial widths in Tab.\,\ref{tab:water-pw} deviate 
more for channels corresponding to high Auger electron energies.

In general, the intensity distribution of our spectra matches the 
experimental spectra well: all peaks and shoulders in the experiments can 
be assigned unequivocally. However, for water, both CBF-CCSD and CBF-MP2 
assign less intensity to the peak with the highest Auger electron energy 
than to the next peak, whereas the opposite pattern was found in the 
experiment. For ammonia, there is no such disagreement, while the two 
respective peaks coalesce for methane. 

The origin of the disagreement in the spectrum of water becomes clear 
by looking at previous theoretical studies. While a different approach 
for the continuum improves the peak intensities only to a certain degree 
as shown in Ref. \citenum{tenorio22}, the consideration of nuclear motion 
via ab initio molecular dynamics (AIMD) as done in Ref. \citenum{inhester12} 
(light blue line in Fig.\,\ref{fig:KLLwater}) resolves most of the 
disagreement with the experiment, both at high energies and low energies. 
We note that the consideration of vibrational effects does not necessarily 
require an AIMD approach. In fact, the impact of vibrational broadening 
on the peak shapes can be captured using a simplified model\cite{cederbaum91} 
that can be combined easily with complex-variable techniques.\cite{camps25} 



\section{K-shell ionization in hydrogen sulfide, phosphine, and silane}
\label{sec:res2}
\subsection{Core ionization energies and total decay widths} \label{sec:res2a}

K-edge core ionization energies for hydrogen sulfide, phosphine, and 
silane are presented in Tab.\,\ref{tab:3rowE}. Compared to the lighter 
molecules (H$_2$O, NH$_3$, CH$_4$), our results for H$_2$S, PH$_3$, and 
SiH$_4$ deviate more from the experimental values. The computed values 
are several eV lower, which can be attributed to scalar relativistic 
effects.\cite{zheng22} However, CBF-CCSD and CBF-MP2 agree with each 
other to the same degree as for the lighter molecules (0.13--0.22\,eV). 
Similar to water (Tab.\,\ref{tab:2rowE}), increasing the number of 
complex-scaled shells has a negligible effect on the core ionization 
energy of H$_2$S. 

\begin{table}[bht]\centering
\caption{K-edge core ionization energies of hydrogen sulfide, phosphine, 
and silane computed with CBF-CCSD and CBF-MP2 in the aug-cc-pCVTZ(5sp) 
basis set with different numbers of complex scaled s-, p-, and d-shells added.
All values in eV.}
\label{tab:3rowE}
\begin{tabular}{lS[table-format=4.2]S[table-format=4.2]S[table-format=4.2]}\hline
Method & {H$_2$S} & {PH$_3$} & {SiH$_4$} \\ \hline
CBF-CCSD$^\text{a}$ & 2472.05 & 2146.20 & 1843.71 \\
CBF-MP2$^\text{a}$ & 2472.18 & 2146.40 & 1843.93 \\
CBF-MP2$^\text{b}$ & 2472.28 & \dots & \dots \\
Expt. & {2478.25\cite{keski74}} & \dots & {1847\cite{bodeur90}} \\ \hline
\end{tabular} \\ \raggedright
$^\text{a}$ 2 complex-scaled s-, p-, and d-shells.\\
$^\text{b}$ 4 complex-scaled s-, p-, and d-shells.\\
\end{table}

Total decay widths are presented in Tab.\,\ref{tab:3rowG}. They are up to 
4 times larger for hydrogen sulfide, phosphine, and silane than for water, 
ammonia, and methane and increase with the charge of the central atom, which 
complies with the expected trend.\cite{agarwal13,mcguire69} For hydrogen 
sulfide, an experimental value is available,\cite{keski74} which agrees 
with our results, taking into account the measurement uncertainty. 

The MP2 total decay widths are 30--70 meV larger than the CCSD values, 
which is a similar absolute deviation as for the second-row compounds 
(Tab.\,\ref{tab:2rowG}). However, it represents a much lower relative 
deviation. Comparing the sum of partial widths with the total width, 
good agreement is only observed for CCSD calculations on H$_2$S and 
PH$_3$, but not for any MP2 calculation. For the 
MP2 calculation on H$_2$S, the addition of more complex-scaled shells 
brings the sum of partial widths closer to the total width.

\begin{table}\centering
\caption{Total decay widths of K-edge core-ionized hydrogen sulfide, phosphine, 
and silane computed with CBF-CCSD and CBF-MP2 in the aug-cc-pCVTZ(5sp) basis set 
with different numbers of complex scaled s-, p-, and d-shells added. All values 
in meV.}
\label{tab:3rowG}
\begin{tabular}{lS[table-format=3.1(3)]S[table-format=3.1]S[table-format=3.1]|S[table-format=3.1]S[table-format=3.1]S[table-format=3.1]}\hline
& \multicolumn{3}{c|}{$\Gamma$} & \multicolumn{3}{c}{$\sum\gamma_{ij}$} \\ \hline
Method & {H$_2$S} & {PH$_3$} & {SiH$_4$} & {H$_2$S} & {PH$_3$} & {SiH$_4$}\\\hline
CBF-CCSD$^\text{a}$ & 443.6 & 423.1 & 343.8 & 446.6 & 416.9 & 386.1 \\
CBF-MP2$^\text{a}$ & 484.1 & 452.0 & 409.7 & 448.1 & 435.6 & 428.8 \\
CBF-MP2$^\text{b}$ & 495.7 & \dots & \dots & 494.9 & \dots & \dots \\
Expt. & {500(100)\cite{keski74}} & \dots & \dots & \dots & \dots & \dots \\ \hline
\end{tabular} \\ \raggedright
$^\text{a}$ 2 complex-scaled s-, p-, and d-shells.\\
$^\text{b}$ 4 complex-scaled s-, p-, and d-shells.\\
\end{table}

\subsection{Partial decay widths} \label{sec:res2b}

Partial widths for H$_2$S, PH$_3$, and SiH$_4$, summarized by the involved 
shells, are presented in Tab.\,\ref{tab:br}. In addition to the absolute values, 
Tab.\,\ref{tab:br} shows the relative distribution among the channels in percent 
as well as a hypothetical distribution based on the assumption that all channels 
have the same width. For H$_2$S, we conducted additional calculations with 4 
instead of 2 complex-scaled s-, p-, and d-shells; these results can be found 
in the supplementary material. The basis-set dependence of these CBF-MP2 widths 
is much less pronounced than for water (Tab.\,\ref{tab:water-pw}). 

\begin{table*}\centering
\caption{Branching ratios for Auger decay of K-edge core-ionized H$_2$S, PH$_3$, 
and SiH$_4$ computed with CBF-CCSD and CBF-MP2 in the aug-cc-pCVTZ(5sp)+2(spd) 
basis.}\label{tab:br}
\begin{tabular}{lrrr|rrr|rrr|rrr|r}\hline
 & \multicolumn{6}{c}{$\sum\gamma_{ij}$ (meV)} & 
 \multicolumn{6}{c}{$\sum\gamma_{ij}/\Gamma$ (\%)} & Same width \\
 & \multicolumn{3}{c}{CCSD} & \multicolumn{3}{c}{MP2} & 
 \multicolumn{3}{c}{CCSD} & \multicolumn{3}{c}{MP2} & for every \\
Branch & H$_2$S & PH$_3$ & SiH$_4$ & H$_2$S & PH$_3$ & SiH$_4$ & H$_2$S & 
PH$_3$ & SiH$_4$ & H$_2$S & PH$_3$ & SiH$_4$ & channel (\%) \\ \hline
L$_1$L$_1$ & 24.6 & 23.4 & 22.8 & 25.8 & 25.0 & 19.8 & 5.5 & 5.6 & 5.9 & 
4.9 & 5.7 & 4.6 & 1.6 \\
L$_1$L$_{2,3}$ & 125.4 & 119.5 & 111.4 & 127.8 & 124.5 & 119.9 & 
28.1 & 28.7 & 28.8 & 27.0 & 28.6 & 28.0 & 9.4 \\
L$_1$M$_1$ & 4.4 & 3.7 & 3.3 & 5.0 & 4.3 & 2.8 & 1.0 & 0.9 & 0.9 & 
0.6 & 1.0 & 0.7 & 3.1 \\
L$_1$M$_{2,3}$ & 7.6 & 5.5 & 3.5 & 7.6 & 5.7 & 3.5 & 1.7 & 1.3 & 0.9 & 
1.7 & 1.3 & 0.8 & 9.4 \\
L$_{2,3}$L$_{2,3}$ & 249.6 & 239.1 & 227.0 & 247.2 & 249.5 & 262.9 & 
55.9 & 57.4 & 58.8 & 57.1 & 57.3 & 61.3 & 14.1 \\
L$_{2,3}$M$_1$ & 8.2 & 6.5 & 5.1 & 8.2 & 6.8 & 5.3 & 1.8 & 1.6 & 1.3 & 
1.9 & 1.6 & 1.2 & 9.4 \\
L$_{2,3}$M$_{2,3}$ & 25.6 & 18.4 & 12.6 & 24.8 & 18.8 & 14.0 & 
5.7 & 4.4 & 3.3 & 6.4 & 4.3 & 3.3 & 28.1 \\
M$_1$M$_1$ & 0.2 & 0.2 & 0.1 & 0.2 & 0.2 & 0.1 & 0.1 & 0.0 & 0.0 & 0.0 & 
0.1 & 0.0 & 1.6 \\
M$_1$M$_{2,3}$ & 0.4 & 0.3 & 0.1 & 0.6 & 0.4 & 0.2 & 0.1 & 0.1 & 0.0 & 
0.1 & 0.1 & 0.0 & 9.4 \\
M$_{2,3}$M$_{2,3}$ & 0.6 & 0.3 & 0.2 & 0.6 & 0.4 & 0.2 & 0.1 & 0.1 & 0.0 & 
0.2 & 0.1 & 0.0 & 14.1 \\ \hline
LL & 399.6 & 382.0 & 361.1 & 400.8 & 399.0 & 402.6 & 89.5 & 91.6 & 93.5 & 
89.0 & 91.6 & 93.9 & 25 \\
LM & 45.8 & 34.1 & 24.5 & 45.6 & 35.7 & 25.7 & 10.2 & 8.2 & 6.4 & 
10.6 & 8.2 & 6.0 & 50 \\
MM & 1.2 & 0.8 & 0.4 & 1.4 & 0.9 & 0.5 & 0.3 & 0.2 & 0.1 & 
0.3 & 0.2 & 0.1 & 25 \\ \hline
\end{tabular} \end{table*}
 
For all molecules, singlet decay channels account for \~\,95\% of the decay 
width, which matches the expectation for K-edge Auger spectra.\cite{siegbahn75,
agren81,matz23a,ferinoperez24} In addition, the largest share of the decay width 
(89\%--94\%) stems from LL channels, while LM channels account for 6\%--11\% and 
the contribution of MM channels remains below 1\%. Thus, assuming equal 
widths for all singlet channels, which is an acceptable approximation for 
Auger decay of second-row elements,\cite{tarantelli87,matz23b,jayadev23} 
is not applicable in the third row of the periodic table. 

The dominance of the LL channels can be related to the energetic and 
spatial proximity of the L shell to the initial hole in the K shell. 
We also note that LL channels contribute somewhat more intensity for 
SiH$_4$ than for PH$_3$ and H$_2$S. In line with the low intensity 
of the LM and MM channels, no measurements of KLM or KMM Auger spectra 
of H$_2$S, PH$_3$, and SiH$_4$ have been reported. Within the KLL 
branch, more than half of the intensity stems from L$_{2,3}$L$_{2,3}$ 
channels, i.e., channels involving two 2p-like orbitals. Their 
dominance is more pronounced for SiH$_4$ than for PH$_3$ and H$_2$S, 
which is in line with experimental results.\cite{vayrynen83} Also, 
we found the same trend for second-row elements in our previous 
work.\cite{matz23a} 

As is evident from Tab.\,\ref{tab:br}, CBF-CCSD and CBF-MP2 predict almost 
identical KLL:KLM:KMM branching ratios. More substantial deviations are 
observed for the KL$_1$L$_1$:KL$_{2,3}$L$_{2,3}$ branching ratios, especially 
for SiH$_4$. This can be traced back to the overestimation of the widths 
involving the 2p orbitals by CBF-MP2. For the widths of individual channels, 
the rms deviation between CBF-CCSD and CBF-MP2 amounts to only 0.5 meV. 
In view of this good agreement, we did not conduct MP2-mod calculations 
according to Eq.\,\eqref{eq:gmp2mod}.

\begin{table} \centering
\caption{Largest partial decay widths of hydrogen sulfide in meV computed 
with CBF-CCSD and CBF-MP2 in the aug-cc-pCVTZ(5sp)+2(spd) basis.}\label{tab:h2s-pw}
\begin{tabular}{@{\extracolsep{8pt}}crr}\hline
Decay & \multicolumn{2}{c}{$\gamma_{ij}$}\\
channel & CCSD & MP2 \\ \hline
$^1$B$_2$ (3a$_1$1b$_2$) & 47.8 & 46.2 \\
$^1$B$_1$ (3a$_1$1b$_1$) & 47.0 & 45.4 \\
$^1$A$_2$ (1b$_1$1b$_2$) & 46.9 & 45.3 \\
$^1$A$_1$ (1b$_1$1b$_1$) & 36.6 & 37.8 \\
$^1$A$_1$ (3a$_1$3a$_1$) & 35.7 & 36.3 \\
$^1$A$_1$ (1b$_2$1b$_2$) & 35.5 & 36.1 \\
$^1$B$_1$ (2a$_1$1b$_1$) & 35.4 & 35.9 \\
$^1$A$_1$ (2a$_1$3a$_1$) & 35.3 & 35.8 \\
$^1$B$_2$ (2a$_1$1b$_2$) & 35.3 & 35.8 \\
$^1$A$_1$ (2a$_1$2a$_1$) & 24.6 & 25.9 \\
$^3$B$_1$ (2a$_1$1b$_1$) & 6.5 & 6.8 \\ \hline
\end{tabular}
\end{table}

The largest individual partial widths for hydrogen sulfide are shown in 
Tab.\,\ref{tab:h2s-pw}. The remaining data are available from the Supporting 
Information. There are several groups of three channels with very similar 
widths, for example, the three strongest channels $^1$B$_2$ (3a$_1$1b$_2$), 
$^1$B$_1$ (3a$_1$1b$_1$), and $^1$A$_2$ (1b$_1$1b$_2$). This similarity is 
a consequence of the atom-like character of the 2p orbitals. In an atom, 
these orbitals would be degenerate and the partial widths would be identical, 
disregarding spin-orbit effects.\cite{jayadev25} 

The small impact of the hydrogen atoms on the 2p orbitals of sulfur in 
H$_2$S also becomes clear by comparing decay widths between H$_2$S and 
H$_2$O. This comparison is shown in the supplementary material. For both 
molecules, the most intense channel is $^1$B$_2$ (3a$_1$1b$_1$), but there 
are significant variations among the remaining channels. The groups of 
three channels that have almost the same width for H$_2$S have different 
widths for H$_2$O. In addition, the $^1$A$_1$ (2a$_1$2a$_1$) channel has a 
higher relative intensity in water than in hydrogen sulfide. 



\subsection{KLL Auger spectra} \label{sec:res2c}

\begin{figure} \centering
\includegraphics[width=\linewidth]{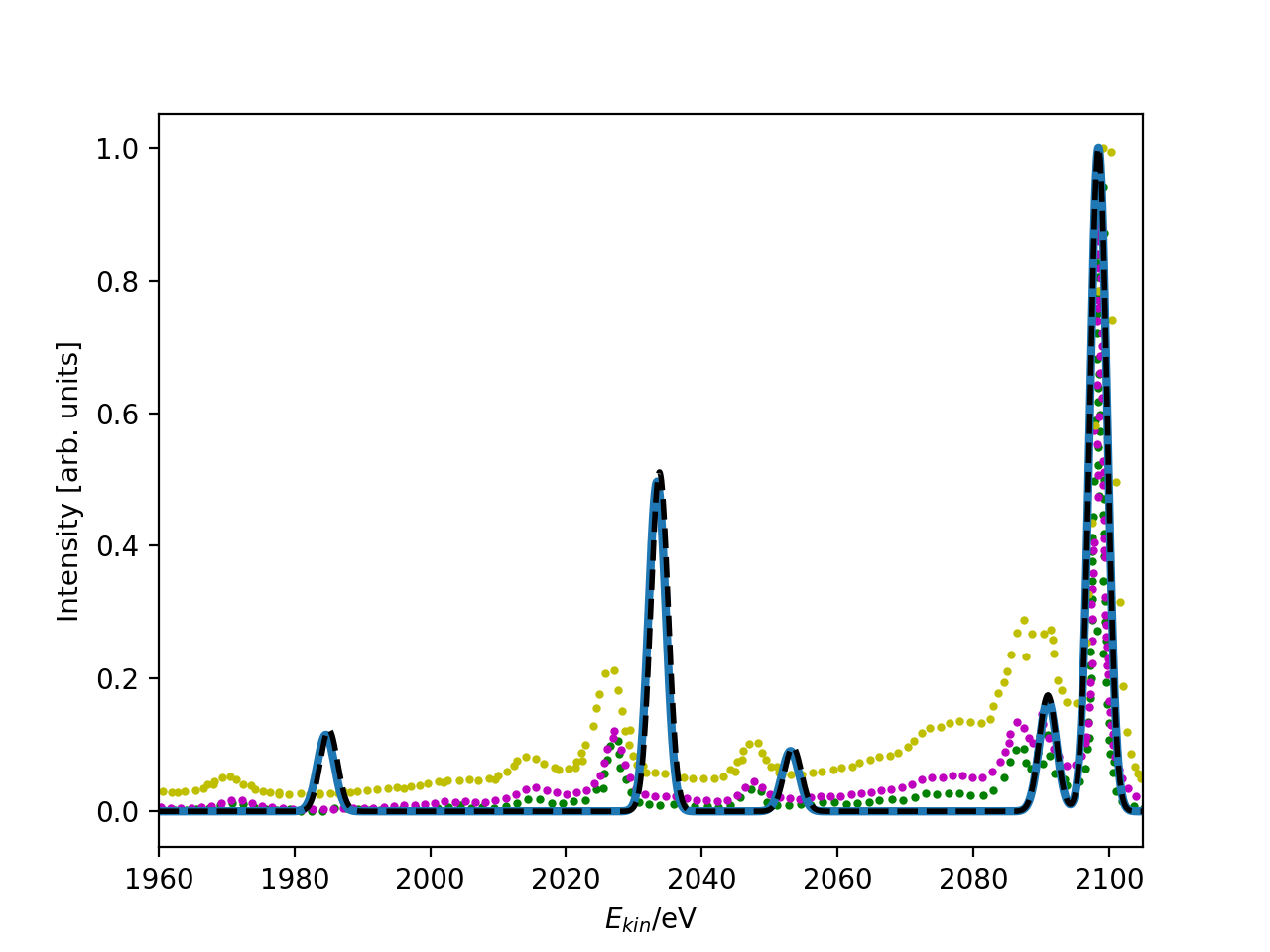}\\
\includegraphics[width=\linewidth]{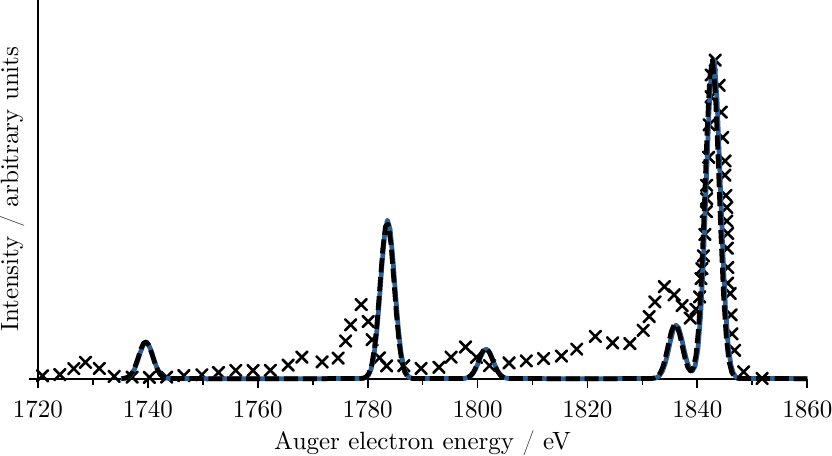}\\
\includegraphics[width=\linewidth]{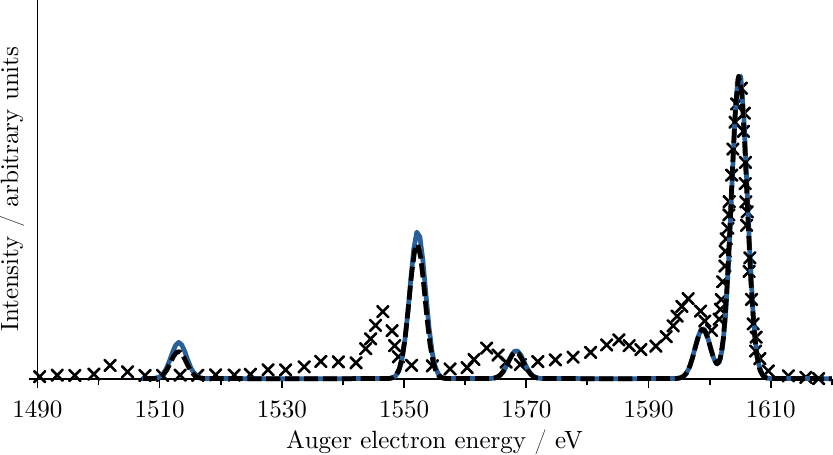}
\caption{KLL Auger spectra of hydrogen sulfide (top), phosphine (middle), 
and silane (bottom). Dark blue full line: CBF-CCSD/EOMDIP-CCSD, black 
dashed line: CBF-MP2/EOMDIP-CCSD(2), green dots: experiment (Ref. 
\citenum{puttner16}), yellow dots: experiment (Ref. \citenum{faegri77}), 
purple dots: experiment (Ref. \citenum{asplund77}), and black $\times$: 
experiment (Ref. \citenum{vayrynen83}).}
\label{fig:h2skll}
\end{figure}

To construct Auger spectra, we computed double ionization energies of 
H$_2$S, PH$_3$, and SiH$_4$ with EOMDIP-CCSD and EOMDIP-CCSD(2). Similar 
to H$_2$O, NH$_3$, and CH$_4$, the deviations between the methods are 
relatively low. For example, for hydrogen sulfide, the maximum deviation 
is 0.50 eV and the rms deviation is 0.38 eV. For LL states, the rms 
deviation is somewhat higher (0.55\,eV) than for LM states (0.25\,eV) 
and MM states (0.12\,eV). This can be rationalized considering that the 
LL double ionization energies are an order of magnitude larger than those 
of MM states. For LL and LM states, EOMDIP-CCSD(2) yields higher double 
ionization energies than EOMDIP-CCSD, while it yields lower values for 
MM states. 

The resulting KLL Auger spectra are shown in Fig.\,\ref{fig:h2skll} together 
with the corresponding experimental spectra.\cite{vayrynen83,asplund77,faegri77,
puttner16} Our computed spectra are shifted to higher energies by 18.35, 17.25, 
and 17.25\,eV (MP2) and 17.85, 17.10, and 17.10 eV (CCSD) for H$_2$S, PH$_3$, 
and SiH$_4$, respectively, so that the highest-energy peak coincides with the 
experimental spectrum. The FWHM is 3.0 eV for all spectra. The intensities of 
the peaks are normalized such that the highest peaks have the same intensity. 

The KLL branch is the only part of the spectrum that has been measured. 
The spectra of H$_2$S, PH$_3$, and SiH$_4$ resemble each other closely; 
7 (PH$_3$, SiH$_4$) and 8 (H$_2$S) peaks can be identified in them. Our 
computed CBF-CCSD and CBF-MP2 spectra are practically indistinguishable 
from each other for every molecule and only comprise five peaks. The 
lowest-energy peak corresponds to the L$_1$L$_1$ (2a$_1$2a$_1$) decay 
channel. The second peak is formed by the three singlet L$_1$L$_{2,3}$ 
states, while the corresponding triplet states form the third peak. The 
fourth peak corresponds to a totally symmetric $^1$A$_1$ transition and 
belongs to the L$_{2,3}$L$_{2,3}$ branch. The highest-energy peak of 
the KLL spectrum is formed by five quasi-degenerate L$_{2,3}$L$_{2,3}$ 
channels.

Compared to the experiment, one observes that the lower-energy signals in 
our computed spectra are shifted to higher energies for all molecules: the 
L$_1$L$_1$ feature by about 14 eV and the L$_1$L$_{2,3}$ signals by about 
7 eV. The likely reason is the use of the truncated EOMDIP operator 
[Eq.\,\eqref{eq:dipmod}] and the subsequent extrapolation procedure 
specified in Sec\,~\ref{sec:th2} for the computation of the double 
ionization energies corresponding to these channels. 

The main disagreement in the peak intensity between experiment and theory 
is that we overestimate the two peaks at lower kinetic energy compared 
to the three highest-energy peaks. We observed a similar overestimation in 
our previous work\cite{matz23b} and could remedy it by taking into account 
the extent of 3-hole-1-particle excitations in the EOMDIP-CCSD wavefunction. 
In the present case, however, this is not possible, as the doubly ionized 
states in question could only be computed with the truncated operator 
from Eq.\,\eqref{eq:dipmod}.

In addition, all experimental spectra have two peaks that are absent in the 
computed spectra: one of them is located at 2015\,eV (H$_2$S), 1768\,eV 
(PH$_3$), and 1537\,eV (SiH$_4$), while the other one is observed at 2075\,eV 
(H$_2$S), 1822\,eV (PH$_3$), and 1585\,eV (SiH$_4$). The nature of these peaks 
remains unclear. Previous theoretical works have related them to shake-up or shake-off 
transitions\cite{krause75,asplund77,pedersen23}, which are not included in our model. In addition, our computed 
spectra do not account for spin-orbit coupling,\cite{jayadev25} which 
may be responsible for the splitting of the peak at 2087 eV in the H$_2$S 
spectrum. 

Comparing H$_2$S, PH$_3$, and SiH$_4$ with H$_2$O, NH$_3$, and CH$_4$, one 
sees that the spectra of the latter molecules span a range of at most 
50\,eV, while those of the former extend over up to 120\,eV. This illustrates 
the larger separation of the orbital energies in the third-row hydrides, 
which is a consequence of the larger ionization energies and ultimately 
the larger nuclear charges. Another difference is the lower number of 
distinct signals in the spectra of the third-row hydrides. This is because 
the atom-like character of the L-shell orbitals renders them almost 
degenerate; the respective decay channels have very similar energies. 
In second-row hydrides, the L-shell orbitals are involved in bonding and 
have different energies depending on their orientation relative to the 
hydrogen atoms.


\subsection{KLM Auger spectra} \label{sec:res2d}

\begin{figure} \centering
\includegraphics[width=\linewidth]{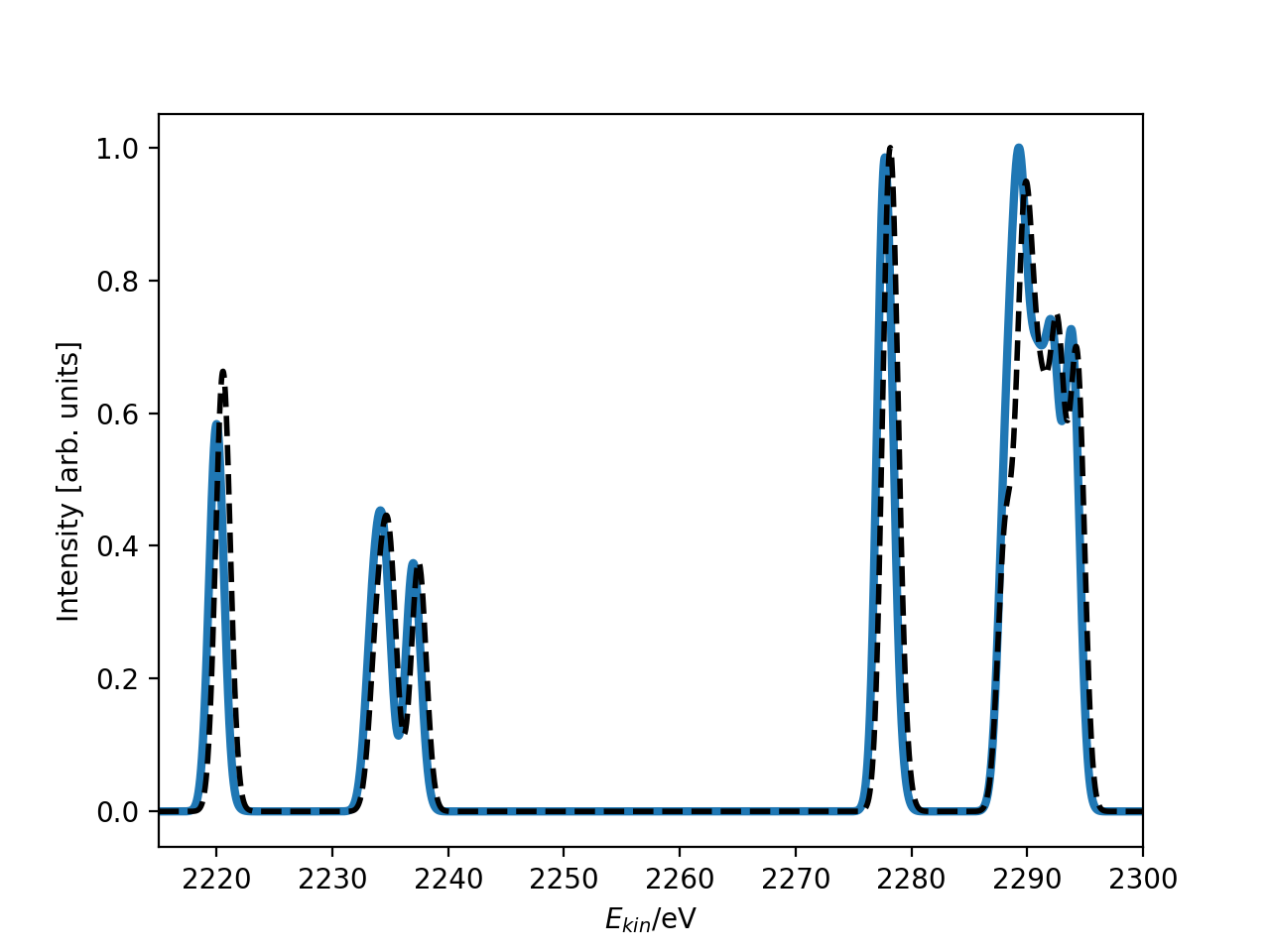}
\caption{KLM Auger spectrum of hydrogen sulfide. Dark blue full line: 
CBF-CCSD/EOMDIP-CCSD, dashed black line: CBF-MP2/EOMDIP-CCSD(2).}
\label{fig:h2sklm}
\end{figure}

Our computed KLM Auger spectra for H$_2$S are shown in Fig.\,\ref{fig:h2sklm}. 
They are shifted by 18.35\,eV (MP2) and 17.85\,eV (CCSD), respectively, to 
higher energies to match the shifts applied in Fig.\,\ref{fig:h2skll}. The 
FWHM is 1.5\,eV for both spectra. 

The KLM spectrum possesses a number of features that is comparable to the KLL spectrum.
7 signals are distinguishable and grouped into two features that are separated 
by a 40 eV gap, which reflects the splitting between the L$_1$ and the L$_{2,3}$ 
orbitals. The higher-energy feature comprises three peaks at 2292\,eV, which are 
so close to each other that they coalesce to a broader signal, and an isolated 
peak at 2278 eV. The 14 eV gap between these two signals corresponds to the 
energy difference between the M$_1$ and M$_{2,3}$ orbitals. The lower-energy 
feature comprises two peaks around 2237 eV and an isolated peak at 2220 eV. The 
gap between these signals again corresponds to the energy difference between 
the M$_1$ and M$_{2,3}$ orbitals. 

The peak positions differ slightly between our two KLM spectra because the 
double ionization energies were shifted for each method separately to align 
the most intense peak in the KLL spectrum. However, it is evident that 
this only leads to local shifts specific to a particular spectral region. 
The peak intensities are in good agreement, with lower-energy signals 
slightly overestimated by CBF-MP2. This is opposite to water, where the 
higher-energy signals were severely overestimated by CBF-MP2. 

\section{L-shell core ionization} \label{sec:res3}

\subsection{Core ionization energies and total decay widths} \label{sec:res3a}

\begin{table}\centering
\caption{L$_1$-edge ionization energies of H$_2$S, PH$_3$, and SiH$_4$ 
computed with CBF-MP2 and CBF-EOMIP-CCSD in the aug-cc-pCVTZ(5sp) basis 
with different numbers of complex-scaled s-, p-, and d-shells added.
All values in eV.} \setlength{\tabcolsep}{7pt}
\begin{tabular}{rrrrr} \hline
Method & CBFs & H$_2$S & PH$_3$ & SiH$_4$ \\\hline
MP2 & +4(spd) & 223.78 & 193.01 & 155.87 \\
MP2 & +5(spd) & 219.53 & 192.99 & 155.86 \\
MP2 & +6(spd) & 229.08 & 192.96 & 155.86 \\
MP2 & +8(spd) & 233.96 & \dots & \dots \\ \hline
EOMIP-CCSD$^\text{a}$ & +6(spd) & 234.98 & 195.36 & 159.04 \\
EOMIP-CCSD$^\text{a}$ & +8(spd) & 234.99 & \dots & \dots \\\hline
Expt. & & 235.0$\pm$0.1$^\text{b}$ & 194.88$^\text{c}$ & 155$^\text{d}$ \\ \hline
\end{tabular}\\ \raggedright
$^\text{a}$ EOMIP-CCSD results for H$_2$S were taken from 
Ref.~\citenum{drennhaus24}. \\
$^\text{b}$ From Ref.~\citenum{hikosaka04}.\\
$^\text{c}$ From Ref.~\citenum{sodhi85}.\\
$^\text{d}$ From Ref.~\citenum{cooper90}.
\label{tab:lshellE}
\end{table}

L$_1$-edge core ionization energies for hydrogen sulfide, phosphine, 
and silane are reported in Tab. \ref{tab:lshellE}. We used EOMIP-CCSD 
as reference method because the CCSD equations could not be converged 
for these states, neither in a complex-scaled basis set nor in an entirely 
real-valued basis set. Although it is no problem to construct the 
respective HF wave functions and compute the MP2 energy using Eq. 
\eqref{eq:emp2}, the large norms of the MP2 amplitude vectors [for example, 
$36-16$i for H$_2$S(2a$_1^{-1}$) as compared to $0.33-0.0007$i for 
H$_2$S(1a$_1^{-1}$), both computed in the aug-cc-pCVTZ (5sp) + 4(spd) 
basis] show that the description is of low quality. 

Indeed, the MP2 L$_1$-edge ionization energies in Tab. \ref{tab:lshellE} 
deviate substantially from EOMIP-CCSD, while we observed a good match 
between MP2 and CCSD for K-edge ionization energies in Tabs. \ref{tab:2rowE} 
and \ref{tab:3rowE}. In addition, the MP2 energy shows a rather erratic 
dependence on the complex-scaled part of the basis set for H$_2$S. 

EOMIP-CCSD agrees with experimental L$_1$-edge ionization energies 
of H$_2$S and PH$_3$ within less than 0.5\,eV. For SiH$_4$, there 
is a deviation of 4 eV, but the experimental value was reported as 
``approximate''.\cite{cooper90}

\begin{table*}\centering
\caption{Total decay widths of L$_1$-edge ionized H$_2$S, PH$_3$, and 
SiH$_4$ computed with CBF-MP2 and CBF-EOMIP-CCSD in the aug-cc-pCVTZ(5sp) 
basis with different numbers of complex-scaled s-, p-, and d-shells added.
All values in meV.} \setlength{\tabcolsep}{7pt}
\begin{tabular}{rrrrrrrr} \hline
Method & CBFs & \multicolumn{2}{c}{H$_2$S} & \multicolumn{2}{c}{PH$_3$} & 
\multicolumn{2}{c}{SiH$_4$} \\
&  & $\Gamma$ & $\sum\gamma_{ij}$ & $\Gamma$ & $\sum\gamma_{ij}$ & 
$\Gamma$ & $\sum\gamma_{ij}$ \\ \hline
MP2 & +2(spd) & 526.3 & 522.2 & 51.3 & 45.2 & 18.9 & 15.1 \\
MP2 & +4(spd) & 3493.9 & 3420.1 & 48.1 & 69.9 & 25.5 & 30.6 \\
MP2 & +5(spd) & 19505.7 & 19524.8 & 77.4 & 94.5 & 26.6 & 37.2 \\
MP2 & +6(spd) & 3708.0 & 3833.1 & 59.0 & 77.8 & 27.8 & 41.1 \\
MP2 & +8(spd) & 1723.4 & 1914.7 & \multicolumn{2}{c}{\dots} & 
\multicolumn{2}{c}{\dots} \\
MP2 & +10(spd) & 99.8 & 106.3 & \multicolumn{2}{c}{\dots} & 
\multicolumn{2}{c}{\dots} \\ \hline
MP2-mod & +5(spd) & {\dots} & 2063.6 & {\dots} & 1676.3 & {\dots} & 1500.2 \\
MP2-mod & +8(spd) & {\dots} & 2157.5 & \multicolumn{2}{c}{\dots} & 
\multicolumn{2}{c}{\dots} \\ \hline
EOMIP-CCSD$^\text{a}$ & +4(spd) & 1119.1 & 1020.0 & 
\multicolumn{2}{c}{\dots} & \multicolumn{2}{c}{\dots} \\
EOMIP-CCSD$^\text{a}$ & +6(spd) & 1603.2 & 1407.4 & 1287.1 & {\dots} & 
1017.8 & {\dots} \\
EOMIP-CCSD$^\text{a}$ & +8(spd) & 1672.2 & 1440.5 & 
\multicolumn{2}{c}{\dots} & \multicolumn{2}{c}{\dots} \\ \hline
Expt.$^\text{b}$ & & \multicolumn{2}{c}{1800} & \multicolumn{2}{c}{\dots} & \multicolumn{2}{c}{\dots} \\ \hline
\end{tabular}\\ \raggedright
$^\text{a}$ EOMIP-CCSD results for H$_2$S were taken from 
Ref.~\citenum{drennhaus24}. \\
$^\text{b}$ From Ref. \citenum{hikosaka04}.
\label{tab:lshellG}
\end{table*}

The total decay widths of the L$_1$-edge ionized states are reported 
in Tab. \ref{tab:lshellG}. These states can undergo L$_1$L$_{2,3}$M 
Coster-Kronig decay, where an electron from the L$_{2,3}$-shell fills 
the vacancy in the L$_1$-shell, and, therefore, have considerably larger 
decay widths than K-edge ionized states.\cite{coster35} This is 
illustrated by the EOMIP-CCSD results in Tab. \ref{tab:lshellG}. 
For H$_2$S, our EOMIP-CCSD decay width is in good agreement with the 
experimental value,\cite{hikosaka04} while no experimental decay widths 
are available for PH$_3$ and SiH$_4$. 

Comparing the sum of EOMIP-CCSD partial widths to the total decay width, 
a larger deviation is observed for the L$_1$-edge ionized states in Tab. 
\ref{tab:lshellG} than for K-edge ionized states (Tabs. \ref{tab:2rowG} 
and \ref{tab:3rowG}). In addition, more complex-scaled shells are needed to 
describe Coster-Kronig decay than Auger decay of K-edge ionized states. 
For a further discussion of these aspects, we refer to our previous 
work on hydrogen sulfide and argon.\cite{drennhaus24}

As can be expected from the analysis in Sec. \ref{sec:thmp2}, the decay 
widths computed with MP2 are completely wrong. For H$_2$S, an erratic 
dependence on the complex-scaled part of the basis set is observed. For 
PH$_3$ and SiH$_4$, results are stable with respect to the basis set 
but too low by more than a factor of 20. The severe underestimation of 
the widths of PH$_3$ and SiH$_4$ can be understood based on Tab. \ref{tab:orb}:
According to the orbital energies of the L$_1$-edge ionized HF wave 
functions, the Coster-Kronig L$_{2,3}$M decay channels are closed. 
The widths in Tab. \ref{tab:lshellG} thus represent only the contribution 
of the MM decay channels, which in reality account for less than 5\%.

To assess the validity of the ``modified'' MP2 method from Sec. 
\ref{sec:thmp2}, we recomputed all decay widths using Eq. \eqref{eq:gmp2mod}. 
The sums of the respective decay widths still deviate by up to 50\% 
from EOMIP-CCSD, but at least they are of the right order of magnitude. In addition, 
the trend in the width going from H$_2$S over PH$_3$ to SiH$_4$ is captured correctly by MP2-mod. 

\subsection{Partial decay widths} \label{sec:res3b}

\begin{table} \centering
\caption{Partial Coster-Kronig decay widths of L$_1$-edge ionized 
hydrogen sulfide computed with CBF-MP2 and ACP-EOMIP-CCSD in the 
aug-cc-pCVTZ(5sp) basis set with different numbers of complex-scaled 
s, p, and d-shells added. All values in meV.} 
\setlength{\tabcolsep}{7pt} 
\begin{tabular}{lrrr} \hline
Decay & EOMIP &  &  \\
channel & -CCSD$^\text{a}$ & MP2-mod$^\text{b}$ & MP2$^\text{c}$ \\ \hline
$^1$B$_1$ (4a$_1^{-1}$1b$_1^{-1}$) & 244.3 & 281.7 & 0.0 \\
$^1$A$_1$ (3a$_1^{-1}$4a$_1^{-1}$) & 236.4 & 288.1 & 0.0 \\
$^1$B$_2$ (4a$_1^{-1}$1b$_2^{-1}$) & 208.7 & 326.3 & 0.0 \\
$^1$A$_1$ (1b$_1^{-1}$2b$_1^{-1}$) & 97.8 & 155.4 & 19060.3 \\
$^1$A$_1$ (3a$_1^{-1}$5a$_1^{-1}$) & 97.2 & 166.0 & 0.1 \\
$^3$A$_1$ (1b$_1^{-1}$2b$_1^{-1}$) & 76.1 & 85.0 & 191.0 \\
$^1$A$_1$ (1b$_2^{-1}$2b$_2^{-1}$) & 65.2 & 111.7 & 0.0 \\
$^3$A$_1$ (1b$_2^{-1}$2b$_2^{-1}$) & 59.8 & 76.4 & 0.0 \\
$^3$A$_1$ (3a$_1^{-1}$5a$_1^{-1}$) & 57.3 & 79.0 & 0.1 \\
$^1$B$_1$ (5a$_1^{-1}$1b$_1^{-1}$) & 43.0 & 73.1 & 0.2 \\
$^3$B$_2$ (3a$_1^{-1}$2b$_2^{-1}$) & 34.1 & 41.6 & --0.1 \\
$^1$B$_2$ (5a$_1^{-1}$1b$_2^{-1}$) & 33.6 & 64.9 & 0.1 \\
$^3$A$_2$ (1b$_2^{-1}$2b$_1^{-1}$) & 32.6 & 36.3 & 60.5 \\
$^3$B$_1$ (3a$_1^{-1}$2b$_1^{-1}$) & 32.2 & 34.4 & 53.8 \\
$^3$A$_2$ (2b$_2^{-1}$1b$_1^{-1}$) & 30.6 & 41.7 & 0.0 \\
$^3$B$_1$ (5a$_1^{-1}$1b$_1^{-1}$) & 26.7 & 35.3 & 0.4 \\
$^3$B$_2$ (5a$_1^{-1}$1b$_2^{-1}$) & 22.9 & 35.0 & --4.5 \\
$^1$B$_2$ (3a$_1^{-1}$2b$_2^{-1}$) & 9.2 & 39.4 & --0.1 \\
$^1$A$_2$ (2b$_2^{-1}$1b$_1^{-1}$) & 8.8 & 39.2 & 0.0 \\
$^1$A$_2$ (1b$_2^{-1}$2b$_1^{-1}$) & 8.3 & 33.0 & 14.7 \\
$^1$B$_1$ (3a$_1^{-1}$2b$_1^{-1}$) & 6.3 & 31.3 & 66.6 \\
$^3$B$_1$ (4a$_1^{-1}$1b$_1^{-1}$) &--5.3 & 3.6 & 0.0 \\
$^3$A$_1$ (3a$_1^{-1}$4a$_1^{-1}$) &--13.7 & 5.9 & 0.0 \\
$^3$B$_2$ (4a$_1^{-1}$1b$_2^{-1}$) &--15.6 & 5.4 & 0.0 \\ \hline
Sum & 1396.5 & 2089.7 & 19443.1\\\hline
\end{tabular} \label{tab:2spw} \\ \raggedright
$^\text{a}$ From Ref. \citenum{drennhaus24}. The basis set is 
aug-cc-pCVTZ(5sp)+8(spd). \\
$^\text{b}$ This work. The basis set is aug-cc-pCVTZ(5sp)+8(spd). \\
$^\text{c}$ This work. The basis set is aug-cc-pCVTZ(5sp)+5(spd).
\end{table}

Partial widths for the L$_{2,3}$M Coster-Kronig decay channels 
of hydrogen sulfide computed with MP2, MP2-mod, and EOMIP-CCSD
are shown in Tab. \ref{tab:2spw}. The corresponding results for 
phosphine and silane and for the MM decay channels can be found 
in the supplementary material. 

The L$_{2,3}$M partial widths computed with the MP2-mod method 
add up to 2090 meV, which is 50\% larger than the EOMIP-CCSD result. 
The width of almost every decay channel is overestimated, but the 
extent of the overestimation varies considerably between 15\% and 70\%. 
However, the dominance of the L$_{2,3}$M decay channels (97\%) over 
the MM decay channels (3\%) is captured correctly and the branching 
ratio between singlet and triplet channels (77\% vs~23\%) is also 
in agreement with EOMIP-CCSD (76\% vs~24\%). We note that there is 
a substantial disagreement between our singlet-triplet branching 
ratios for hydrogen sulfide and those computed with multiconfigurational 
Dirac-Hartree-Fock theory for the argon atom (45\% vs~55\%),\cite{liu21} 
which has a similar electronic structure.

With the conventional MP2 method, the majority of the partial widths 
in Tab. \ref{tab:2spw} are completely unphysical. Many open channels,
including those for which the largest widths are expected, have zero 
decay width. This has the same reason as the wrong total widths of 
PH$_3$ and SiH$_4$ discussed in the previous section, namely, the 
orbital energies of the L$_1$-edge ionized HF wave function (see 
Tab. \ref{tab:orb}).

At the same time, the widths of a few other channels are overestimated 
by MP2. In particular, the $^1$A$_1$ (1b$_1$2b$_1$) channel has a 
width of 19060 meV, which is responsible for the erroneous total 
width in Tab. \ref{tab:lshellG}. We examined this channel further 
by analyzing the contributions of individual virtual orbitals to Eq. 
\eqref{eq:gmp2}. Most of the 19060 meV can be traced back to one 
single summand where the denominator in Eq. \eqref{eq:gmp2} assumes 
a value of $-0.0005 - 0.0008\text{i}$ a.u., i.e., almost zero. 

The respective orbital energies are in atomic units 
$\varepsilon_{1\text{b}_1^\beta} = -7.5166+0.0000008\text{i}$, 
$\varepsilon_{2\text{b}_1^\alpha} = -0.8261-0.000004\text{i}$, 
$\varepsilon(\text{core hole})_{2\text{a}_1^\beta} = 
-8.2654+0.000006\text{i}$, and $\varepsilon_{8\text{a}_1^\alpha} = 
-0.0768-0.000844\text{i}$.

The occurrence of such near-singularities depends on the virtual 
orbital energies, which in turn strongly depend on the basis set. 
For PH$_3$ and SiH$_4$, no near-singularities occur, so that the 
basis-set dependence of the respective MP2 results in Tab. 
\ref{tab:lshellG} remains inconspicuous.  

\subsection{LLM Coster-Kronig spectra} \label{sec:res3c}

\begin{figure} \centering
\includegraphics[width=\linewidth]{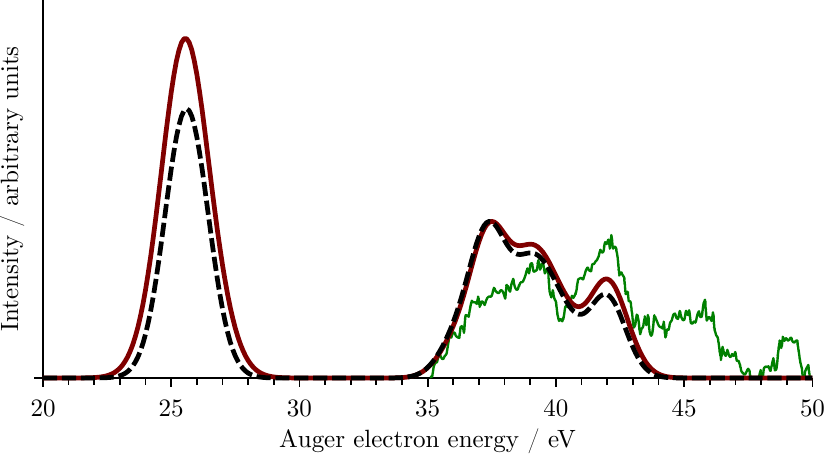} \\
\includegraphics[width=\linewidth]{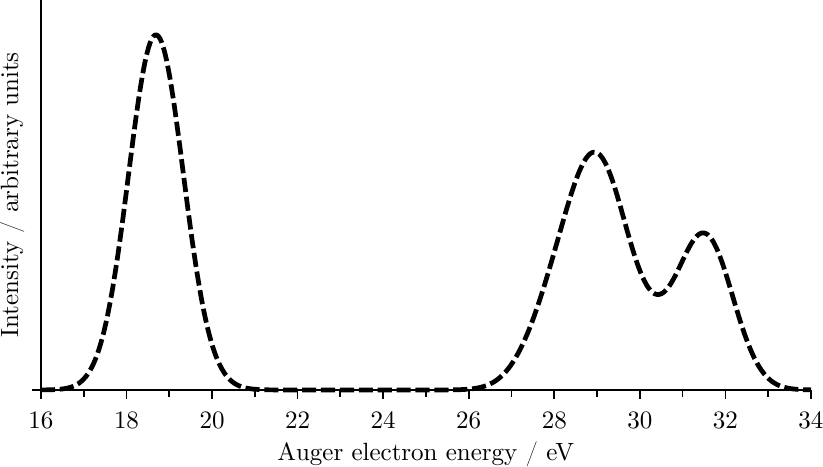} \\
\includegraphics[width=\linewidth]{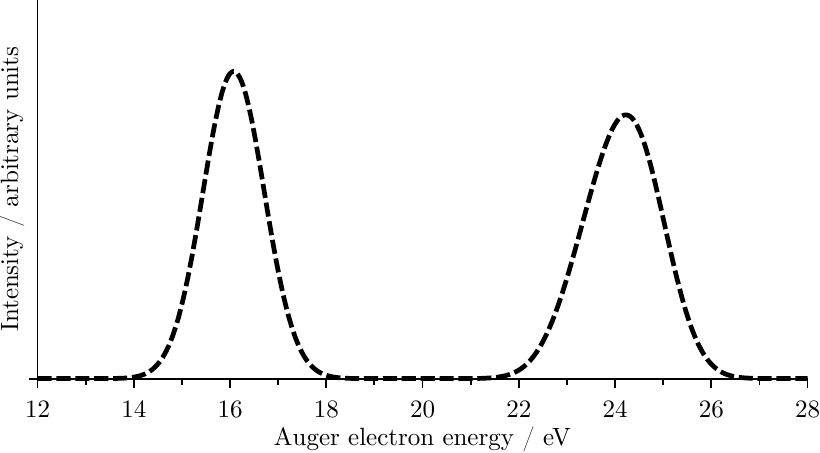}
\caption{Coster-Kronig spectra of hydrogen sulfide (top), phosphine 
(middle), and silane (bottom). Black dashed line: CBF-MP2-mod/EOMDIP-CCSD(2), 
red line: ACP-EOMIP-CCSD/EOMDIP-CCSD (Ref.~\citenum{drennhaus24}), green line: 
experiment(Ref.~\citenum{hikosaka04}).}
\label{fig:costerkronig}
\end{figure}

The LLM Coster-Kronig spectra for hydrogen sulfide, phosphine, and 
silane computed with the MP2-mod method are shown in Fig. 
\ref{fig:costerkronig}. The FWHM is 1.5\,eV for all spectra, and for 
H$_2$S, our spectrum is shifted to higher energies by 2.9\,eV. Hydrogen 
sulfide is the only molecule for which an experimental spectrum is 
available.\cite{hikosaka04} In addition, we previously computed this 
spectrum with EOMIP-CCSD.\cite{drennhaus24}

The spectrum for H$_2$S illustrates good agreement between MP2-mod 
and EOMIP-CCSD, even though the corresponding partial widths (Tab. \ref{tab:2spw}) 
deviate substantially. This is because the branching ratios are in better agreement than the absolute values of the decay widths. The relative intensity of the signal around 
25\,eV, which is formed by the L$_{2,3}$M$_1$ decay channels, is 
slightly underestimated with MP2-mod, but the feature between 35 
and 45\,eV, which is formed by the L$_{2,3}$M$_{2,3}$ decay channels, 
is obtained with a shape that is almost identical to the EOMIP-CCSD 
spectrum. 

The experimental spectrum is incomplete and only covers the region 
above 35\,eV, where it shows two major signals, a broader one 
between 37 and 39\,eV and a sharper one at 42\,eV. While this is 
in line with our computed spectra, the relative intensities are not 
in agreement. In addition, the experimental spectrum includes some further 
signals with weaker intensity at higher energies (45-50 eV) that 
the computed spectra lack. We tentatively assign them to resonant 
decay of core-excited states. 

Note that we suggested a different assignment in our previous work 
(Ref. \citenum{drennhaus24}). There, we assigned the peaks with the 
highest intensity in the experimental spectrum and the computed 
spectrum to each other. With this alternative assignment, there 
are no unassigned signals at higher energy. However, a large shift 
of the computed spectrum by 7.5\,eV was required to match the 
experimental spectrum and the feature below 40 eV remained unassigned. 
Therefore, we believe that the current assignment as shown in Fig. 
\ref{fig:costerkronig}, is more sensible. 

As concerns phosphine and silane, Fig. \ref{fig:costerkronig} shows 
that the emitted Coster-Kronig electrons have lower energies than 
for hydrogen sulfide, while the spectra have a similar shape. For 
all three molecules, the lower-energy peak formed by the L$_{2,3}$M$_1$ 
decay channels is more intense than the higher-energy signals 
corresponding to the L$_{2,3}$M$_{2,3}$ decay channels. The spacing 
between the two signals shrinks from 12\,eV (H$_2$S) over 11 eV 
(PH$_3$) to 8 eV (SiH$_4$), reflecting the decreasing energy 
difference between the 3s (M$_1$) and 3p (M$_{2,3}$) orbitals. 
In addition, the higher-energy signal consists of one peak in SiH$_4$, 
while there are two peaks in PH$_3$ and three in H$_2$S. This 
reflects the degeneracy of the highest occupied molecular 
orbital in SiH$_4$ (t$_2$), whereas there are two and three 
corresponding energy levels in PH$_3$ and H$_2$S, respectively.

\section{Conclusions} \label{sec:con}


In this work, we presented an approach for the computation of Auger 
and Coster-Kronig spectra in the framework of MP2 theory combined 
with complex-scaled basis functions (CBFs). Based on our earlier 
work about complex-scaled CCSD,\cite{matz22} we showed how total 
and partial decay widths can be computed from the complex-valued MP2 
energy of core-ionized states.

We applied this approach to the decay of K and L$_1$-ionized states of 
H$_2$O, NH$_3$, CH$_4$, H$_2$S, PH$_3$, and SiH$_4$. For K-edge decay 
of H$_2$S, PH$_3$, and SiH$_4$, CBF-MP2 works very well. The dominance of 
the LL decay channels (>90\%) over LM (1\%--10\%) and MM (<1\%) decay 
channels is captured correctly, and the rms deviation of the individual 
partial decay widths from CBF-CCSD amounts to only 0.5 meV. 

For the corresponding decay processes in H$_2$O, NH$_3$, and CH$_4$, 
CBF-MP2 performs significantly worse, with the rms deviation of the 
individual partial decay widths from CBF-CCSD amounting to 4 meV. The 
widths of some channels are overestimated by 50\%, resulting in a too 
large total width and distorted branching ratios. 

For Coster-Kronig decay of L$_1$-ionized states of H$_2$S, PH$_3$, and 
SiH$_4$, CBF-MP2 fails completely. We traced this back to the fact that 
L$_{2,3}$M decay channels are closed in the HF wave functions of 
L$_1$-ionized states. By using orbital energies from the HF wave 
functions of the corresponding neutral states in the MP2 energy expression, 
qualitatively correct results can be obtained for the decay of L$_1$-ionized 
states. Moreover, the description of the decay of K-ionized states of 
H$_2$O, NH$_3$, and CH$_4$ is improved somewhat as well. 

To describe the final states of Auger decay, we introduced in this work 
a double ionization potential variant of the EOM-CCSD(2) method,\cite{stanton95} 
also known as EOM-MP2. The double ionization energies computed with this 
method show an rms deviation from EOMDIP-CCSD that amounts to 0.08 eV for 
H$_2$O, NH$_3$, and CH$_4$ and to 0.38 eV for H$_2$S, PH$_3$, and SiH$_4$,  
which is acceptable for most purposes in the context of Auger spectroscopy. 

Taken together, the construction of Auger spectra from CBF-MP2 and 
EOMDIP-CCSD(2) calculations entails substantially lower computational 
cost than our previous approach based on CBF-CCSD and EOMDIP-CCSD without 
compromising accuracy. At the same time, the need to shift the orbital 
energy denominator for some states limits the viability of the CBF-MP2 
method. This shortcoming is avoided in an EOM-CC approach in which all 
states relevant to Auger decay are built from the same reference wave 
function that is optimized for the neutral molecule.\cite{matz23a}

The comparison of the Auger spectra computed in this work to experimental 
data shows overall good agreement regarding peak positions and intensities 
but also illustrates the impact of effects that are not included in our 
current theoretical approach. For example, the impact of nuclear motion 
is rather prominent in the Auger spectrum of water, and spin-orbit effects 
are visible for hydrogen sulfide. We also hope that the LLM Coster-Kronig 
spectra of PH$_3$ and SiH$_4$ and the KLM spectrum of H$_2$S presented in 
our work foster measurements of these spectra.

\section*{Supplementary Material}
The supplementary material is available online and contains explicit expressions for the EOMDIP-CCSD(2) $\sigma$ vectors, details about the used complex-scaled basis functions and optimal scaling angles as well as plots showing the energy extrapolation procedure discussed in section II C. Furthermore, it includes detailed numerical data on partial decay widths for all examined molecules with all used basis sets and a list of all the channels with their configurations, energies, and decay widths that were used to generate the Auger spectra. Finally, an Auger spectrum using a variation of the MP2-mod approach is presented and compared to the spectrum generated according to Eq.~\ref{eq:gmp2mod}.

\section*{Acknowledgments}
T.-C. J. gratefully acknowledges funding from the European Research 
Council (ERC) under the European Union’s Horizon 2020 research and 
innovation program (Grant Agreement No. 851766) and the KU Leuven 
internal funds (Grant No. C14/22/083). The authors are grateful to 
Professor Sonia Coriani for helpful comments regarding the errors in the 
core ionization energies due to relativistic effects.

\section*{Data Availability Statement}
The data that support the findings of this study are available within the article and its supplementary material.


\section*{References}
\bibliographystyle{achemso}
\bibliography{aipsamp}

\end{document}